\documentclass[11pt,a4paper]{article}
\pdfoutput=1

\usepackage{jheppub}
\usepackage{subfigure}
\usepackage{epstopdf}
\usepackage[T1]{fontenc}
\usepackage{extarrows}

\title{Nonlinear chiral transport from holography}
\author[a]{Yanyan Bu,}
\author[b]{Tuna Demircik,}
\author[b]{and Michael Lublinsky}
\affiliation[a]{Department of Physics, Harbin Institute of Technology, Harbin 150001, China}
\affiliation[b]{Department of Physics, Ben-Gurion University of the Negev,
Beer-Sheva 84105, Israel}

\emailAdd{yybu@hit.edu.cn}
\emailAdd{demircik@post.bgu.ac.il}
\emailAdd{lublinm@bgu.ac.il}

\abstract{Nonlinear transport phenomena induced by the chiral anomaly are explored within a 4D field theory defined holographically as $U(1)_V\times U(1)_A$ Maxwell-Chern-Simons theory in Schwarzschild-$AdS_5$. First, in presence of external electromagnetic fields,  a general form of vector and axial currents is derived. Then, within the gradient expansion up to third order, we analytically compute all (over 50) transport coefficients.
A wealth of higher order (nonlinear) transport phenomena induced by chiral anomaly are found beyond the Chiral Magnetic and Chiral Separation Effects.
Some of the higher order  terms are relaxation time corrections to the lowest order nonlinear effects.
The charge diffusion constant and dispersion relation of the Chiral Magnetic Wave are found to receive  anomaly-induced non-linear corrections due to e/m background fields.
Furthermore, there emerges a new gapless mode, which we refer to as {\it Chiral Hall Density Wave},  propagating along the background Poynting vector.}


\arxivnumber{1807.08467}
\begin{document}
\maketitle

\flushbottom

\allowdisplaybreaks

\section{Introduction}\label{intro}
Hydrodynamics \cite{landau,forster} is an effective low energy description of many interacting QFTs near thermal equilibrium. Historically, hydrodynamics has been always associated with a long wavelength  limit  of the underlying microscopic theory, while over the last decade or so there is an increased number of works addressing ``hydronization'' relaxing  the long wavelength approximation. Rather, hydrodynamics is defined as an effective  theory of conserved currents, such as stress tensor and/or charge currents, assuming their algebra is closed on a relevant set of near-equilibrium states.

Dynamics of the theory is governed by conservation equations (continuity equations) of the currents. The simplest example
is $\partial_t\rho=-\vec\nabla\cdot\vec J$, which is a time evolution equation for the charge density $\rho$ sourced by three-current $\vec J$.
However, this equation cannot be solved as an initial value problem without additional input, the current $\vec J$.  In hydrodynamics,
$\vec J$ has to be expressed in terms of thermodynamical variables, such as $\rho$ itself, temperature, and possibly external fields if present.
This is known as constitutive relation.
Traditionally, in the long wavelength limit,  constitutive relations are presented as a (truncated) gradient expansion.
At any given order, this expansion is fixed by thermodynamic considerations and symmetries, up to a finite number of transport coefficients (TCs).
The latter should be  either computed from underlying microscopic theory  or deduced experimentally. Diffusion constant,
DC conductivity or shear viscosity are examples of the lowest order TCs.

It is well known, however,  that in relativistic theory  truncation of the gradient expansion at any fixed order
leads to serious conceptual problems
such as violation of causality.
Beyond conceptual issues, causality violation results in numerical instabilities rendering the entire framework unreliable.
Causality is restored  when all order gradient terms are included, in a way providing a UV completion to the ``old'' hydrodynamic effective theory.
Below we will refer to such case as \textit{all order resummed} hydrodynamics \cite{Lublinsky:2007mm,0905.4069,1406.7222,1409.3095,1502.08044,1504.01370}. The first completion of the type
was originally proposed by  M\"{u}ller, Israel, and Stewart (MIS) \cite{Muller1967,ISRAEL1976310,ISRAEL1976213,ISRAEL1979341} who  introduced retardation effects in the constitutive relations for the currents. The MIS formulation \cite{Muller1967,ISRAEL1976310,ISRAEL1976213,ISRAEL1979341}
is the most popular scheme  employed  in practical simulations. Recent ideas  on the nature of the hydrodynamic expansion, gradient resummation and attractor behavior, etc. could be found in \cite{0905.4069,1302.0697,1503.07514,1507.02461,1609.02820,1704.08699,
 Spalinski:2017mel,1710.03234,1707.02282,Denicol:2018pak,Behtash:2017wqg,Behtash:2018moe}.

In this paper we continue exploring  hydrodynamic regime of  relativistic plasma with chiral asymmetries. We closely follow previous works \cite{1608.08595,1609.09054}  focusing on massless fermion plasma with two Maxwell gauge fields, $U(1)_V\times U(1)_A$.
As a result  of  chiral anomaly,  which appears in relativistic QFTs with massless fermions,
global $U(1)_A$ current  coupled to external electromagnetic fields is no longer conserved.
The continuity equations turn into
\begin{equation}\label{cont eqn}
\partial_{\mu}J^{\mu}=0, \qquad \qquad \partial_{\mu}J_5^{\mu}=12 \kappa \vec{E}\cdot \vec{B},
\end{equation}
where $J^{\mu}/J_5^{\mu}$ are vector/axial currents and $\kappa$ is an anomaly coefficient ($\kappa=eN_c/(24\pi^2)$ for $SU(N_c)$ gauge theory with a massless Dirac fermion in fundamental representation and $e$ is  electric charge, which will be set to unit from now on). $\vec{E}$ and $\vec{B}$ are vector electromagnetic fields.
Non-conservation of the axial current in (\ref{cont eqn}) receives extra contribution if external axial electromagnetic fields are turned on.
Throughout this work, however, we will not consider external axial fields (they were considered in Ref \cite{1608.08595}). Chiral plasma plays a major role in a number of fundamental research areas, historically starting from primordial plasma in the
early universe \cite{Kuzmin:1985mm,Vilenkin:1982pn,Rubakov:1996vz,Grasso:2000wj,doi:10.1142/S0218271804004530}. During the last decade,
macroscopic effects induced by the chiral anomaly were found to be of relevance in relativistic heavy ion collisions \cite{025420,Kharzeev:2013ffa,Huang:2015oca}, and have been searched intensively at LHC \cite{1512.05739,1610.00263,1708.01602,1708.08901,Skokov:2016yrj}.
Finally, (pseudo-)relativistic systems in condensed matter physics, such as Dirac and
Weyl semimetals, display anomaly-induced phenomena, which were recently observed experimentally \cite{Liu:2014,Lv:2015pya,Xu:2015cga,2014ARCMP,1412.6543,1503.01304,1507.06470} and can be studied via similar theoretical methods \cite{1207.5808,Landsteiner:2014vua,Jimenez-Alba:2015awa,Landsteiner:2015lsa}.

A hydrodynamic description of (chiral) plasma amounts to  solving  a set of coupled equations.  As has been mentioned earlier, the continuity equations
({\ref{cont eqn}}) have to be supplemented by constitutive relations describing plasma medium effects. Generically, these are of the type
\begin{equation} \label{const}
\vec J\ =\ \vec J\;[\rho, \rho_5, T, \vec E, \vec B]; \hspace{3cm}  \vec J_5=\vec J_5[\rho, \rho_5, T, \vec E, \vec B],
\end{equation}
where $\rho_5$ is the axial charge density and $T$ stands for the temperature\footnote{ We  prefer to parameterise the currents (\ref{const}) in terms of the charge densities $\rho,\rho_5$
because it is more natural and straightforward within the holographic framework. Yet, we could switch to a more traditional representation with the chemical potentials $\mu,\mu_5$ as hydrodynamical variables
(see Section \ref{Main results} for details).}.

In a sense, the constitutive relations (\ref{const}) are ``off-shell'' relations, because they treat the charge
density $\rho$ ($\rho_5$) as independent of $\vec J$  ($\vec J_5$). Employing (\ref{cont eqn}), the currents (\ref{const}) are put into ``on-shell''.
In (\ref{const}), the fields $\vec E,\vec B$ are assumed to be {\it external}. However, the charges and currents  induce e/m fields of their own. Thus, the external electromagnetic fields $\vec E,\vec B$ have to be promoted into dynamical ones, satisfying  Maxwell equations (in Gaussian units)
\footnote{In principle, the axial sources $(\rho_5,\vec J_5)$, through another set of chiral anomaly-modified Maxwell's equations,  would also generate classical axial e/m fields.
In their turn, the axial e/m fields would enter and modify the constitutive relations (\ref{const}), see e.g. \cite{1608.08595}.},
\begin{align}
\vec\nabla \cdot \vec E= 4\pi \rho^{\rm tot},&\qquad \qquad \vec\nabla \times \vec B=\frac{1}{c}\left(4\pi {\vec J}^{\rm \;tot}+\partial_t \vec E\right), \label{dynamical Maxwell bdry}\\
\vec \nabla \cdot \vec B=0,&\qquad \qquad \vec \nabla \times \vec E=-\partial_t \vec B, \label{Bianchi bdry}
\end{align}
where $\rho^{\rm tot}$ and ${\vec J}^{\rm \;tot}$ are the total charge density and total current, a sum of external sources  $(\rho^{\rm ext}$,$\vec J^{\rm \;ext})$ and
induced part $(\rho,\vec J)$, which is the one that enters the  constitutive relations (\ref{const}). The external sources could be absent  when
a fully isolated system is considered. A typical example would be primordial plasma in the early Universe frequently studied using
magneto-hydrodynamics (MHD).  MHD, along with many other effective theories of the type,   also involves  neutral flow dynamics. That is, in addition to the charge current sector
discussed above, one has to simultaneously consider energy-momentum conservation. Generically, the two dynamical sectors are coupled. However, in the discussion below, we will consider the {\it probe limit}, under which one ignores back-reaction of the charge sector on the energy-momentum conservation. This implies $\varepsilon+p\gg \mu \rho +\mu_5 \rho_5$ with $\varepsilon,p$ being the fluid's energy density and pressure.

A self-consistent evolution of the system is determined by solving together (\ref{cont eqn},\ref{const}, \ref{dynamical Maxwell bdry}) given some initial conditions.  While the equations (\ref{cont eqn}, \ref{dynamical Maxwell bdry}) are exact, the constitutive relations  (\ref{const}) are the ones where various hydrodynamic approximations are applied.   A great deal of modelling
normally enters (\ref{const}),  such as truncated gradient expansion, weak field approximation, etc.  As a result of a full simulation, one sometimes finds instabilities leading
to exponential growths of some quantities, such as  of dynamical magnetic fields.  It thus becomes mandatory to check if the original approximations made for the constitutive relations
are consistent with the solutions found. If not,  the hydrodynamical model has to be revised.

We just outlined a general setup for a hydrodynamical problem, but it is not our goal here to carry it over for any realistic system. Instead, motivated by the discussion above we would like to focus on the nature of the constitutive relations (\ref{const}), which are well known to receive contributions induced by the chiral anomaly.
The most familiar example is the \textit{chiral magnetic effect} (CME) \cite{PhysRevD.22.3080,0808.3382}:  a vector current is generated along an external magnetic field  when  a chiral imbalance between left- and right-handed fermions is present ($\vec J\sim \rho_5\vec B$).  There is a vast literature on CME, which we cannot review here in full. The chiral magnetic conductivity was computed in perturbative QCD in \cite{Kharzeev:2009pj,Fukushima:2010vw,1103.2035,Satow:2014lva,Yee:2014dxa,Jimenez-Alba:2015bia}. In \cite{hep-ph/0511236,Yee:2009vw,Rebhan:2009vc,Matsuo:2009xn,Gorsky:2010xu,Rubakov:2010qi,Gynther:2010ed,1102.4577,
Kalaydzhyan:2011vx,1106.4030,Hu:2011ze,Hu:2011qa,Bai:2012ci,Lin:2013sga,1506.01384,Grozdanov:2016ala,1701.05565}
it was evaluated for the strong coupling regime using  AdS/CFT correspondence \cite{Maldacena:1997re,Gubser:1998bc,Witten:1998qj}.
CME emerged via arguments based on the second law of thermodynamics, that is positivity of entropy production \cite{0906.5044,Sadofyev:2010pr}, and also within the  chiral kinetic theory (CKT) \cite{1203.2697,1207.0747,1210.8158,1203.0725,1210.8312}.
Finally,  numerical evidence based on lattice gauge theory for CME can be found in
\cite{Buividovich:2009wi,Abramczyk:2009gb,Fukushima:2009ft,Braguta:2010ej,Yamamoto:2011gk,Braguta:2013loa}.
We would like to comment by passing that CME is believed to be a strict non-equilibrium phenomenon. In other words, different arguments indicate that CME must  vanish in equilibrium \cite{1207.5808,Kharzeev:2013ffa,1502.01547,1605.08724}
\footnote{We thank Mikhail Zubkov for bringing this issue to our attention. We also thank Dmitri Kharzeev, Shu Lin, Andrey Sadofyev, and Ho-Ung Yee for stimulating discussions about this point.}.

Another important transport phenomenon induced by the chiral anomaly is the \textit{chiral separation effect} (CSE) \cite{Son:2004tq,Metlitski:2005pr}:
left and right charges get separated along  applied external magnetic field ($\vec J_5\sim \vec B$).
Combined,  CME and CSE lead to a new gapless excitation called \textit{chiral magnetic wave} (CMW) \cite{Kharzeev:2010gd}. This is
a propagating wave along the magnetic field. While signature of CME/CSE has not yet been confirmed in heavy ion collision experiments \cite{1512.05739,1610.00263,1708.01602,1708.08901},  a large negative longitudinal magneto-resistance observed in Dirac/Weyl semimetals can be attributed to CME \cite{1412.6543,1503.01304,1507.06470}.

Just like in Refs. \cite{1608.08595,1609.09054}, our playground will be a holographic model, namely $U(1)_V\times U(1)_A$ Maxwell-Chern-Simons theory in Schwarzschild-$AdS_5$ \cite{Yee:2009vw,Gynther:2010ed}
to be introduced in detail in Section \ref{s2}.
%, for which we know how to compute a zoo of transport coefficients exactly.
For some sort of universality, we hope to learn from this model about both generic structures of the currents and relative strengths of various effects.

Recently, transport phenomena {\it nonlinear} in external fields were  realised \cite{Avdoshkin:2014gpa} to be of critical importance in having self-consistent evolution of chiral plasma.
Combined with the causality arguments mentioned earlier, the conclusion is that the constitutive relations (\ref{const}) should contain some ``nonlinear'' transport coefficients so to
guarantee their applicability in a  broader regime.
Particularly, traditional MHD is strongly affected by anomalous transports \cite{Joyce:1997uy,Boyarsky:2011uy,Manuel:2015zpa,Boyarsky:2015faa,Hirono:2015rla},
which necessitates a development of a fully self-consistent chiral MHD.
This triggered strong interest in nonlinear chiral transport phenomena within CKT  \cite{1603.03620,1603.03442,1705.01267,1710.00278},
 to which we will compare some of our findings below. Previous works on the subject of nonlinear anomalous transports
include \cite{1105.6360} based on the entropy current approach and  \cite{1304.5529} based on the fluid-gravity correspondence.

The main objective of the series of publications \cite{1608.08595,1609.09054,BDL} and the present work is to explore  the constitutive relations (\ref{const})
under various approximations, primarily zooming  on  transport phenomena induced by the chiral anomaly. In the present publication, the following new directions are explored.
First, we derive  general  expressions for the vector and axial currents, see (\ref{formal current1}, \ref{formal current2}), which do not involve any approximations.  This clarifies the concept of
``non-renormalisation'' of CME/CSE \cite{Gursoy:2014boa,Gursoy:2014ela,1608.08595,1609.09054} when  electromagnetic fields can be both strong and inhomogeneous in spacetime. Second,within the holographic model, we complete the calculation of all second order nonlinear transport coefficients and  compare  with those obtained in CKT \cite{1603.03442}. Finally, and this is the main novel part  in this publication, all third order transport coefficients  are computed analytically, including relaxation time corrections to some second order transport terms (See Section (\ref{const})).
This paves a way for the gradient resummation project released in \cite{BDL}: some of the third order transport coefficients become all order frequency/momentum-dependent  functions.

In the next Section, we will review our results including  connections to the previous works \cite{1608.08595,1609.09054} and the forthcoming publication \cite{BDL}.
The remaining Sections present details of  the calculations.

\section{Summary of the results}

\subsection{Generalities}

This subsection briefly summarises  the series of works \cite{1608.08595,1609.09054,1511.08789,BDL} including the present one, so to
help the reader to navigate between various studies and results. We write down the most comprehensive  constitutive relation and indicate specific approximations applied in each individual work.

Following  \cite{1608.08595,1609.09054},
the charge densities and external fields are split into constant backgrounds and space-time dependent fluctuations
\begin{equation} \label{linsch1}
\begin{split}
&\rho(x_\alpha)=\bar{\rho}+\epsilon\delta\rho(x_\alpha), \qquad\qquad \rho_5(x_\alpha)=\bar{\rho}_5+\epsilon\delta\rho_5(x_\alpha),\\
&\vec{E}(x_\alpha)=\vec{\bf{E}}+\epsilon\delta\vec{E}(x_\alpha), \qquad\quad \vec{B}(x_\alpha)=\vec{\bf{B}}+\epsilon\delta\vec{B}(x_\alpha),
\end{split}
\end{equation}
where $\bar{\rho}$, $\bar{\rho}_5$, $\vec{\bf{E}}$ and $\vec{\bf{B}}$ are the backgrounds, while
$\delta\rho$, $\delta\rho_5$, $\delta\vec{E}$ and $\delta\vec{B}$ stand for the fluctuations. Here
$\epsilon$ is a formal expansion parameter to be used below. Furthermore, being unable to perform calculations for arbitrary background fields for most of the time, we introduce an expansion in the field strengths
\begin{equation} \label{linsch2}
\vec{\bf{E}}\rightarrow\alpha\vec{\bf{E}}, \quad\quad \vec{\bf{B}}\rightarrow\alpha\vec{\bf{B}},
\end{equation}
where $\alpha$ is a corresponding  expansion parameter. Below we will introduce yet another expansion parameter $\lambda$,
which will correspond to the hydrodynamical gradient expansion. For the purpose of gradient counting, e/m fields will be frequently considered as  $\mathcal{O}(\lambda^1)$.

%Any term in the currents' constitutive relations takes a form:
The constitutive relations (\ref{const}) can be formally Taylor expanded in all its arguments. This  includes the gradient ($\lambda$),  $\epsilon$, and $\alpha$ expansions.
Parametrically, a generic term entering (\ref{const}) looks like
\begin{equation} \label{generic gradient}
\bar{\rho}^{\;k}\;\bar{\rho}_5^{\;k_5}\;\vec{\bf{E}}^{\;n_E}\;\vec{\bf{B}}^{\;n_B}\;
\partial_t^{\;m_t}\;\vec{\nabla}^{\;m_x}\left(\delta\rho^{\;l}\;\delta\rho_5^{\;l_5}\;
\delta\vec{E}^{\;l_E}\;\delta\vec{B}^{\;l_B}\right),
\end{equation}
which is multiplied by a transport coefficient\footnote{In fact, each term in (\ref{generic gradient}) corresponds to a large number of terms obtained by different actions of the derivatives and index contractions.}. $k$, $k_5$, $n_E$, $n_B$, $m_t$, $m_x$, $l$, $l_5$, $l_E$, $l_B$ are integers.
%Here we assumed Taylor expansion for the constitutive relations.
The most general constitutive relations correspond to a sum of all possible terms like (\ref{generic gradient})\footnote{The asymptotic nature of the gradient expansion and problems related to resummation of the series have been a hot topic over the last few years, see recent works \cite{1302.0697,1503.07514,1509.05046}. In our approach, however, we never attempt to actually sum the series and thus these discussions
are of no relevance to our formalism.}.

Obviously, we do not intend to consider all possible terms in (\ref{generic gradient}). Instead, most of  the results obtained in the present and  early works \cite{1608.08595,1609.09054}  can be combined in a compact constitutive relation
(focusing on the vector current $\vec{J}$\;),
\begin{align} \label{jmu collection}
\vec{J}=&\gamma_1\vec\nabla \rho+ \gamma_2\vec\nabla\rho_5+ \gamma_3\vec{E}+ \gamma_4 (\rho_5 \vec{B})+ \gamma_5\vec\nabla \times \vec{B} + \gamma_6 (\vec{E}\times \vec\nabla\rho) + \gamma_7 \vec{B}\times (\rho\vec\nabla\rho)\nonumber\\
+&\gamma_{18} \vec{B}\times (\rho_5\vec\nabla\rho_5)  + \gamma_8 (\vec{E}\times \vec\nabla\rho_5) + \gamma_9 (\rho \vec E \times \vec B) + \gamma_{10} \vec{\nabla} \left(\vec{B}\cdot \vec{\nabla}\rho_5\right) + \gamma_{11} \vec{\nabla} \left(\vec{B} \cdot \vec{\nabla}\rho\right)\nonumber\\
+& \gamma_{12}(\rho \vec\nabla B^2) + \gamma_{13} (\rho \vec B)+ \gamma_{14} \vec\nabla(\vec E\cdot \vec\nabla \rho)+\gamma_{15}(\rho\vec E) +\gamma_{16} \vec\nabla(\vec E\cdot \vec\nabla \rho_5)+\gamma_{17} (\rho_5\vec E).
\end{align}
The  coefficients $\gamma_i$ are most general $O(3)$ scalars which could be constructed from three vectors $\vec\nabla$, $\vec E$, and $\vec B$.
That is,  $\gamma_i$ are scalar  functions of $E^2$ and $B^2$, and pseudo-scalar functions of $\vec E\cdot \vec B$. Furthermore,
$\gamma_i$ are  scalar functionals of derivative operators $\partial_t$, $\vec\nabla^2$, $\vec E \cdot\vec\nabla$,  and pseudo-scalar functionals of $\vec B \cdot\vec\nabla$,
\begin{equation}
\gamma_i=\gamma_i\left(\partial_t,\vec\nabla^2,\vec E \cdot\vec\nabla,\vec B \cdot\vec\nabla;E^2,B^2,\vec E\cdot \vec B\right),
\end{equation}
which correspond to all order gradient resummation, as mentioned in Introduction. $\gamma_i$ themselves are rich in structure and  contain information about non-linear
corrections in the fields.
Taylor expanding $\gamma_i$  in all their arguments (all the derivatives are assumed to act on the right of $\gamma_i$) gives rise to each individual gradient term like (\ref{generic gradient}).
Admittedly, (\ref{jmu collection})  does not contain all the possible terms like in (\ref{generic gradient}).  Particularly,
while the constitutive relation (\ref{jmu collection}) does contain some  nonlinear in $\rho,\rho_5$ terms,
it excludes most of the nonlinear terms of the third order, which are collected in (\ref{nonlinrhorho5j},\ref{nonlinrhorho5j5}). Some of the terms in (\ref{jmu collection})  are well recognisable, such as diffusion ($\gamma_1$), electrical conductivity ($\gamma_3$), or  CME ($\gamma_4$).
Some other terms might be less familiar and we will discuss them below in detail.

As explained in the Introduction, the purpose of \cite{1608.08595,1609.09054,BDL} and the present   work  is to systematically explore (\ref{generic gradient}) under different approximations.
%To make various studies and results reported in \cite{1608.08595,1609.09054} and here more easily accessible,
We first briefly summarise  them
using both the notations of (\ref{generic gradient}) and (\ref{jmu collection}), and then deepen our presentation of the current study.

$\bullet$  Ref \cite{1608.08595}, study  1.
No background fields, $\vec {\bf E}=\vec{\bf B}=0$;
all order gradient terms that are linear in the inhomogeneous fluctuations $\delta\rho,\delta \rho_5,\delta \vec E,\delta \vec B$ are resummed\footnote{In \cite{1608.08595} we also considered transports related to axial external electromagnetic fields.}. This corresponds to calculating currents up to  $\mathcal{O}(\epsilon^1\alpha^0)$.
%As the second study of \cite{1608.08595}, we focused on nonlinear corrections to vector/axial currents due to the constant pieces $\vec {\bf E},\vec{\bf B}$. By powers of $\epsilon,\alpha$, this second study derives currents up to order $\mathcal{O}(\epsilon^0\alpha^3)$.
Using the notations (\ref{generic gradient}) and (\ref{jmu collection}) this study corresponds to
%constitutive relations for different sections of \cite{1608.08595} could be classified as
\begin{eqnarray}
&&\underline{\text{\cite{1608.08595}-1}_a}:\quad n_E=n_B=0,\; l+l_5+l_E+l_B=1, \;\forall (k,k_5), \;m_t+m_x\leq3\;\Rightarrow \;\text{(analytic)} \nonumber \\
&&\underline{\text{\cite{1608.08595}-1}_n}:\quad n_E=n_B=0,\; l+l_5+l_E+l_B=1, \;\forall (k,k_5),\; \forall (m_t,m_x)\;\Rightarrow \;\text{(numeric)} \nonumber \\
&&\quad\gamma_i=\gamma_i(\partial_t,\vec\nabla,0,0;0,0,0),\qquad i=1,3,4, 5.
%&&\underline{\text{I3}}:\quad n_E+n_B\leq3, \; l=l_5=l_E=l_B=m_t=m_x=0, \;\forall(k,k_5)\;\Rightarrow \;\text{(analytic)}.  \nonumber
\end{eqnarray}
%In accord with (\ref{jmu collection}), in the first study of \cite{1608.08595}---{\rm I1 \& I2} we computed
%\begin{equation}
%\gamma_i=\gamma_i(\partial_t,\vec\nabla,0,0;0,0,0),\qquad i=1,3,4, 5,
%\end{equation}
%but with the rest missed.
The remaining $\gamma_i$ have not been probed in the study.
$\gamma_i(\partial_t,\vec\nabla)$ correspond to the  gradient resummation. Thanks to the linearisation, the constitutive relations could be conveniently  expressed in
Fourier space. Then, the functionals of the derivatives are turned into functions of frequency and space momenta, $(\partial_t, \vec \nabla) \to (-i\omega,i\vec{q}) $.
We refer to $\gamma_i(-i\omega, q^2) $ as transport coefficients functions (TCFs) \cite{1409.3095}. TCFs contain information about infinitely many derivatives and associated transport coefficients. In practice, TCFs are not computed as a series resummation of order-by-order  hydrodynamic expansion, and are in fact exact to all orders.  TCFs go beyond the hydrodynamic low frequency/momentum limit and contain collective effects of non-hydrodynamic modes. Fourier transformed back into real space, TCFs become memory
functions. Diffusion and shear viscosity memory functions were previously computed in \cite{1511.08789,1502.08044}.

$\bullet$  Ref. \cite{1608.08595}, study  2.
Nonlinear in $\vec {\bf E}$ and $\vec{\bf B}$ corrections to the vector/axial currents. The currents are derived up to  $\mathcal{O}(\epsilon^0\alpha^3)$.
%In the second study of \cite{1608.08595}---{\rm I3}, only the following transports are taken into account
\begin{eqnarray}
%&&\underline{\text{I1}}:\quad n_E=n_B=0,\; l+l_5+l_E+l_B=1, \;\forall (k,k_5), \;m_t+m_x\leq3\;\Rightarrow \;\text{(analytic)} \nonumber \\
%&&\underline{\text{I2}}:\quad n_E=n_B=0,\; l+l_5+l_E+l_B=1, \;\forall (k,k_5),\; \forall (m_t,m_x)\;\Rightarrow \;\text{(numeric)} \nonumber \\
&&\underline{\text{\cite{1608.08595}-2}}:\quad n_E+n_B\leq3, \; l=l_5=l_E=l_B=m_t=m_x=0, \;\forall(k,k_5)\;\Rightarrow \;\text{(analytic)} \nonumber \\
&&\quad \gamma_i=\gamma_i(0,0,0,0;E^2,B^2,\vec E\cdot \vec B),\qquad i=3,4,9.
\end{eqnarray}
%\begin{equation}
%\gamma_i=\gamma_i(0,0,0,0;E^2,B^2,\vec E\cdot \vec B),\qquad i=3,4,11.
%\end{equation}

$\bullet$ Ref. \cite{1609.09054}, study  1.
Nonlinear corrections to vector/axial currents due to  static but spatially-inhomogeneous magnetic field.
\begin{eqnarray}
&&\underline{\text{\cite{1609.09054}-1}_a}:\quad n_E=l_E=0,\;
 \forall (l,l_5,k,k_5),\; n_B+m_t+m_x+l_B\leq2\; \Rightarrow\;\text{(analytic)} \nonumber\\
&&\underline{\text{\cite{1609.09054}-1}_a}:\quad n_E=l_E=0,\;
 \forall (l,l_5,k,k_5),\; n_B+m_t+m_x+l_B=3, \;l+l_5+k+k_5=1\nonumber\\
&& \qquad\qquad \quad \Rightarrow\;\text{(analytic)} \nonumber\\
% &&\underline{\text{II3}}:\quad n_E=l_B=l=k=k_5=m_x=0,\;
 %\forall (n_B,l_5,l_E),\; m_t+n_B+l_E\leq3, \; \Rightarrow\;\text{(analytic)} \nonumber\\
 %&&\underline{\text{II4}}: \quad n_E=l_B=l=k=k_5=m_x=0,\; l_E+l_5=1,\;\forall n_B,\; \forall m_t \Rightarrow \text{(numeric)}\nonumber.
%&&\quad  \gamma_i=\gamma_i(\partial_t,\vec\nabla,0,\vec{B}\cdot \vec\nabla;0,B^2,0),\qquad i=1,4,5,7,10,11,12.\\
&& \gamma_{1,4}=\gamma_{1,4}(\partial_t,0,0,0;E^2,0,0),\qquad \gamma_{5}= \gamma_{5}(0,0,0,0;0,0,0), \nonumber\\
&&\gamma_{7,18}= \gamma_{7,18}(0,0,0,0;0,0,0),\qquad \gamma_{10,11,12}= \gamma_{10,11,12}(0,0,0,0;0,0,0). \nonumber
\end{eqnarray}

$\bullet$ Ref. \cite{1609.09054}, study  2. Dependence of longitudinal electric conductivity on arbitrary strong constant magnetic field.
A time-varying electric field is assumed to be weak.
%Since the spatial dependence was turned off in the second study of \cite{1609.09054}---{\rm II3 \& II4}, we only considered transports
\begin{eqnarray}
%&&\underline{\text{II1}}:\quad n_E=l_E=0,\;
 %\forall (l,l_5,k,k_5),\; n_b+m_t+m_x+l_B\leq2\; \Rightarrow\;\text{(analytic)} \nonumber\\%
%&&\underline{\text{II2}}:\quad n_E=l_E=0,\;
% \forall (l,l_5,k,k_5),\; n_b+m_t+m_x+l_B=3, \;l+l_5+k+k_5=1\; \Rightarrow\;\text{(analytic)} \nonumber\\
 &&\underline{\text{\cite{1609.09054}-2}_a}:\quad n_E=l_B=l=k=k_5=m_x=0,\;
 \forall (n_B,l_5,l_E),\; m_t+n_B+l_E\leq3, \; \Rightarrow\;\text{(analytic)} \nonumber\\
 &&\underline{\text{\cite{1609.09054}-2}_n}: \quad n_E=l_B=l=k=k_5=m_x=0,\; l_E+l_5=1,\;\forall n_B,\; \forall m_t \Rightarrow \text{(numeric)}\nonumber.\\
&&\quad \gamma_3=\gamma_3(\partial_t,\vec{\nabla},0,0;0,0,\vec E\cdot \vec B), \qquad \gamma_4=\gamma_4(\partial_t,\vec{\nabla},0,0;E^2,0,0)
\end{eqnarray}

$\bullet$ In the present work,
 we relax some of the approximations made in  \cite{1608.08595,1609.09054}  and derive constitutive relations for the currents, up to third order in the gradient expansion.
 %assuming slowly-varying hydrodynamic variables and external fields (which or some of which are assumed to be constant background in \cite{1608.08595,1609.09054}).
 %Secondly, regarding TCFs, we extend the analysis of \cite{1608.08595} to the order $\mathcal{O}(\epsilon^1\alpha^1)$ and systematically resum nonlinear transports. Using the notation (\ref{generic gradient}), the present study could be sketched as
\begin{eqnarray}
%&&n_E+n_B+l_E+l_B+m_t+m_x\leq2,\;
% \forall (l,l_5,k,k_5), \Rightarrow\;\text{(analytic)} \nonumber\\
%&& n_E+n_B+l_E+l_B+m_t+m_x=3,\; l+l_5+k+k_5=1 \Rightarrow\;\text{(analytic)} \nonumber\\
&& n_E+n_B+l_E+l_B+m_t+m_x\leq3,\; \forall (l,l_5,k,k_5) \Rightarrow\;\text{(analytic)} \nonumber\\
% &&\underline{\text{III3}}:\quad l_E=l_B=0,\; n_E+n_B=1, \;l+l_5=1, \; \forall (k, k_5), \; m_t+m_x\leq3\; \Rightarrow\;\text{(analytic)} \nonumber\\
% &&\underline{\text{III4}}: \quad l_E=l_B=0,\; n_E+n_B=1, \;l+l_5=1, \; \forall (k, k_5,m_t,m_x)\; \Rightarrow\; \text{(numeric)}\nonumber.
&& \gamma_{1,4}=\gamma_{1,4}(\partial_t,0,0,0;E^2,0,0),\qquad \gamma_2=\gamma_2(0,0,0,\vec{B}\cdot \vec\nabla;0,0,0), \nonumber\\
&& \gamma_3=\gamma_3(\partial_t,\vec\nabla,0,0;0,0,\vec{E}\cdot \vec{B}),\qquad
\gamma_{5,8,9}= \gamma_{5,8,9}(\partial_t,0,0,0;0,0,0), \nonumber\\
&&\gamma_{7,18}= \gamma_{7,18}(\partial_t,0,0,0;0,0,0),\qquad \gamma_{10,11,12}= \gamma_{10,11,12}(0,0,0,0;0,0,0).
\end{eqnarray}

$\bullet$ In the forthcoming paper \cite{BDL}, we will primarily focus on those TCFs associated with nonlinear terms at $\mathcal{O}(\epsilon^1\alpha^1)$.
%{\it ???????The analysis of \cite{1608.08595} is extended to the order $\mathcal{O}(\epsilon^1\alpha^1)$ and systematically resum nonlinear transports.}
\begin{eqnarray}
%&&\underline{\text{III1}}:\quad n_E+n_B+l_E+l_B+m_t+m_x\leq2,\;
 %\forall (l,l_5,k,k_5), \Rightarrow\;\text{(analytic)} \nonumber\\
%&&\underline{\text{III2}}:\quad n_E+n_B+l_E+l_B+m_t+m_x=3,\; l+l_5+k+k_5=1 \Rightarrow\;\text{(analytic)} \nonumber\\
 &&\underline{\text{\cite{BDL}-1}_a}:\quad l_E=l_B=0,\; n_E+n_B=1, \;l+l_5=1, \; \forall (k, k_5), \; m_t+m_x\leq3\; \Rightarrow\;\text{(analytic)} \nonumber\\
 &&\underline{\text{\cite{BDL}-1}_n}: \quad l_E=l_B=0,\; n_E+n_B=1, \;l+l_5=1, \; \forall (k, k_5,m_t,m_x)\; \Rightarrow\; \text{(numeric)}\\
 && \gamma_{1,2}=\gamma_{1,2}(\partial_t,\vec\nabla,\vec E\cdot \vec\nabla,\vec B\cdot \vec \nabla;0,0,0),\quad \gamma_i=\gamma_i(\partial_t,\vec\nabla,0,0;0,0,0),\quad i=4,6-8,15,17,18.\nonumber
\end{eqnarray}
In the next subsection, we summarise the main results of the present work,
leaving all the technical details in the main text and Appendix.

\subsection{Main results}\label{Main results}

This subsection is further split into three parts. The first one contains the general form of the currents.
The second part focuses  on the second order nonlinear transport and comparison with similar results obtained in the CKT.
The third order terms constituting the bulk of our new results  are presented towards the end of this subsection.

\subsubsection{General form of the currents
%: CME non-renormalised
}

A formal expression for the constitutive relations for the vector and axial currents is derived in Section \ref{s3}, having the following form
% charge densities and external electromagnetic fields are generic.
%In compact form, the currents' constitutive relations are
\begin{eqnarray}
\label{formal current1}
&&J^t=\rho,~~~~~~~~~~~~\vec{J}=-\mathcal{D}_0\vec{\nabla} \rho+\sigma_e^0\vec{E}+ \sigma_{\chi}^0\mu_5 \vec{B}+ \delta \vec{J},\\
\label{formal current2}
&&J^t_5=\rho_5,~~~~~~~~~~~\vec{J}_5=-\mathcal{D}_0\vec{\nabla} \rho_5+\sigma_{\chi}^0\mu \vec{B}+ \delta \vec{J}_5,
\end{eqnarray}
where $\rho,\,\rho_5$ are {\it generic} vector/axial charge densities and $\mu,\,\mu_5$ are corresponding
chemical potentials. The external e/m fields $\vec E, \vec B$ are  {\it generic} as well, that is  none of the approximations introduced in \cite{1608.08595,1609.09054} is assumed.
The lowest order TCs---charge diffusion constant $\mathcal{D}_0$, DC electrical conductivity $\sigma_e^0$ and DC CME/CSE conductivity $\sigma_\chi^0$ are \cite{1511.08789,1608.08595}
\begin{equation}
\mathcal{D}_0=\frac{1}{2},\qquad \sigma_e^0=1,\qquad \sigma_\chi^0=12\kappa,
\end{equation}
where from here on we set $\pi T=1$ for convenience. Proper powers of $\pi T$ for dimensionfull quantities could be easily recovered given their physical dimensions.

The $\sigma_\chi^0$-terms in $\vec{J}$ and $\vec{J}_5$ are standard CME and CSE, respectively, in agreement with  ``non-renormalisability"  of CME \cite{1012.6026,Gursoy:2014ela,Gursoy:2014boa}.
%Here, we conclude this feature of CME for an arbitrary magnetic field $\vec B(t,\vec x)$,
%particularly relaxing the approximations undertaken in \cite{1608.08595, 1609.09054, Gursoy:2014boa, Gursoy:2014ela}.
The new element here, which we find  important to emphasise,  is that  (\ref{formal current1}, \ref{formal current2}) are exact, that is they are derived relaxing all the approximations undertaken
in \cite{1608.08595, 1609.09054, Gursoy:2014boa, Gursoy:2014ela}.
%in the sense that they are derived without any approximation for $\rho$, $\rho_5$, $\vec{E}$ and $\vec{B}$.
Nonlinearity of CME/CSE in external fields $\vec E$ and $\vec{B}$ is completely absorbed into the chemical potentials $\mu$, $\mu_5$.

The corrections $\delta \vec{J}$ and $\delta \vec{J}_5$ are formally defined in (\ref{deltaJJ5}), which consist of higher derivative terms starting from the second order {\it only} .
These terms are  built from powers and derivatives of  $\vec{E},\vec{B}$, $\mu$ and $\mu_5$.  $\delta \vec{J}$ and $\delta \vec{J}_5$ are not known analytically, but could be worked
out perturbatively.  $\delta \vec{J}$ and $\delta \vec{J}_5$
introduce new effects,  particularly additional contributions to the currents along the direction of $\vec{B}$ proportional to derivatives of the chemical potentials.
These effects introduce very important modifications to the original CME/CSE. As will be clear later, external e/m fields make corrections to $\mathcal{D}_0$ and $\sigma_e^0$, and even generalise them into tensor-type TCs.
While in principle an axial analogue of $\sigma_e^0$ (i.e., a term proportional to  $\vec{E}$-term) in $\vec{J}_5$ is also possible, it does not appear in our calculations due to the probe limit \cite{1803.08389}.

We have mentioned earlier a discussion about vanishing equilibrium CME, which might appear in tension with (\ref{formal current1}).
In principle, since $U(1)_A$ is not a symmetry,  axial gauge potential
$\mathcal{A}_\mu$ itself could be regarded as another external field. Our calculations, however, are performed assuming vanishing  $\mathcal{A}_\mu$.
Had we introduced a non-vanishing constant background for the time component,  $\mathcal{A}_t\neq 0$, CME conductivity $\sigma_\chi^0$ would be shifted
\begin{equation} \label{CME shift}
\vec{J}_{\rm CME}=12\kappa (\bar \mu_5- \mathcal{A}_t) \vec{B},
\end{equation}
due to a Chern-Simons contribution \cite{Gynther:2010ed,1608.08595}.  In order to have CME vanish, it suffices to impose $\mathcal{A}_t=\bar{\mu}_5$  \cite{Gynther:2010ed,1207.5808}.
%In \cite{Gynther:2010ed,1207.5808} the authors took the requirement $\mathcal{A}_t=\bar{\mu}_5$ as a prescription to obtain a vanishing CME current for in-equilibrium state
In \cite{1207.5808} it was indeed argued that the equality $\mathcal{A}_t=\bar{\mu}_5$ must be satisfied in equilibrium.
While we do not have much to add to this discussion, we notice that
%We, however, have no conclusive remark about this point.
a constant $\mathcal{A}_t$ does not lead to  any new  effect beyond the shift (\ref{CME shift}) in CME. This is because  the bulk dynamics underlying our model is expressed
entirely in terms of the vector and axial field strengths.
%and thus will not introduce further modifications beyond the shift of DC limit of CME (\ref{CME shift}). Therefore, our results are still quite general.

\subsubsection{Second order results: comparison with the CKT}

At second order, the results read
\begin{equation} \label{delJ 2nd}
\begin{split}
\delta \vec{J}=&\frac{1}{4}\sigma_m^0 \left(\rho^2+\rho_5^2\right) \vec{\nabla} \times\vec{B} -\frac{1}{4}\mathcal{D}_H^0\vec{B}\times\left(\rho\vec{\nabla} \rho+\rho_5\vec{\nabla} \rho_5\right)-\frac{1}{2}\sigma_{a\chi H}^0 \vec{E}\times \vec{\nabla} \rho_5 \\
&-\tau_e\sigma_e^0 \partial_t \vec{E}-\frac{1}{2}\sigma_{\chi H}^0\;  \rho \vec{B}\times \vec{E} -\frac{1}{2}\tau_\chi  \rho_5 \partial_t \vec{B} +\tau_D\partial_t \vec{\nabla} \rho+\tau_{\bar\chi} (\partial_t\rho_5) \vec{B}+ \mathcal{O}(\partial^3),
\end{split}
\end{equation}
\begin{equation} \label{delJ5 2nd}
\begin{split}
\delta \vec{J}_5=&\frac{1}{2}\sigma_m^0 \rho\rho_5 \vec{\nabla}\times \vec{B} -\frac{1}{4} \mathcal{D}_H^0 \vec{B}\times \left(\rho \vec{\nabla} \rho_5+ \rho_5 \vec{\nabla} \rho\right)-\frac{1}{2}\sigma_{a\chi H}^0 \vec{E}\times \vec{\nabla}\rho \\
&-\frac{1}{2}\sigma_{\chi H}^0  \rho_5 \vec{B}\times\vec{E}-\frac{1}{2}\tau_\chi \rho \partial_t \vec{B}+\tau_D \partial_t\vec{\nabla} \rho_5 +\tau_{\bar\chi} (\partial_t\rho) \vec{B}+ \mathcal{O}(\partial^3),
\end{split}
\end{equation}
where the TCs take the following values
\begin{equation}\label{t1}
\sigma_m^0=72(2\log2-1) \kappa^2, \qquad \qquad \mathcal{D}_H^0=72(3\log2-2) \kappa^2, \qquad \qquad \sigma_{a\chi H}^0=6\log2\,\kappa,
\end{equation}
\begin{equation}
\sigma_{\chi H}^0=72\log2\, \kappa^2,\qquad \qquad  \tau_e=\frac{\log2}{2}, \qquad \qquad \tau_\chi=12\log2\,\kappa,
\end{equation}
\begin{equation}\label{t8}
\tau_D= \frac{\pi}{8}, \qquad \qquad \tau_{\bar\chi}= - \left( \frac{3}{2} \pi +3\log2 \right) \kappa.
\end{equation}
%In (\ref{t1}-\ref{t8}) we have used the convention $\pi T=1$, which will be used throughout subsequent presentations.
The TCs in (\ref{delJ 2nd},\ref{delJ5 2nd}) could be related to Taylor expansion of some $\gamma_i$'s in (\ref{jmu collection}), schematically indicated as follows
\begin{equation} \label{tau1-8}
\begin{split}
&\sigma_m^0\in \gamma_5,\qquad \mathcal{D}_H^0 \in \gamma_8, \qquad \sigma_{a\chi H}^0 \in \gamma_9, \qquad \sigma_{\chi H}^0 \in \gamma_{11},\qquad \tau_e \in \gamma_3,\\
&\tau_\chi, \tau_{\bar \chi} \in \gamma_4,\qquad \tau_D \in \gamma_1.
\end{split}
\end{equation}
%{\it The constitutive relations (\ref{delJ 2nd},\ref{delJ5 2nd}) are an extension of the ones obtained in \cite{1608.08595,1609.09054}. While most of the terms and respective coefficients
%are not new (except for $\sigma_{a\chi H}^0$, which is indeed new),
%the results are shown to have a broader range of applicability as, in contrast to \cite{1608.08595,1609.09054},  they do not rely on any
%assumptions of $\rho,\rho_5,\vec{E},\vec{B}$ being  static or homogeneous.
%}
%\\
%{\color{red} The first referee complained about this point, since he seemed to be confused about which contents are new. I think about this issue a lot, but ended without a good rephrasing. We may just remove this paragraph completely, or just add one comment "via including spacetime variations of e/m fields (i.e., $\delta \vec E$ and $\delta\vec B$), the instability issue (linearly growing of $\rho_5$ with respect to time) of the axial continuity equation is removed."}
%
To the best of our knowledge, in a holographic model, $\sigma_{a\chi H}^0$ is computed here for the first time. The rest of the TCs in (\ref{delJ 2nd}, \ref{delJ5 2nd}) have been previously computed in \cite{1608.08595,1609.09054}. While in the constitutive relations these terms were already introduced in \cite{1608.08595,1609.09054}, they did not contribute to dynamics
(continuity equations) due to the static/homogeneous field approximations assumed in these earlier publications.
%More precisely, under various approximations of \cite{1608.08595,1609.09054}, only a sub-sector of terms in (\ref{delJ 2nd}, \ref{delJ5 2nd}) could be covered.
For instance, to $\mathcal{O}(\epsilon^0)$  considered in \cite{1608.08595}, the currents do not contain any gradient terms, resulting in dynamical instability (linear growth of $\rho_5$
with time).
%(as reflected in the axial continuity equation).
Inclusion  of the external e/m field fluctuations (i.e., $\delta \vec E$ and $\delta\vec B$) and associated gradient terms in (\ref{delJ 2nd}, \ref{delJ5 2nd}) regularizes
the instability.  Our results (\ref{delJ 2nd}, \ref{delJ5 2nd}) reveal new dynamical effects and thus  are novel in a sense of a much  broader range of applicability.
Physical interpretation  of the terms in (\ref{delJ 2nd}, \ref{delJ5 2nd}) and  their dissipative properties are discussed below.

%In (\ref{delJ 2nd},\ref{delJ5 2nd}) we introduced four types of relaxation time terms: $\tau_e$ for electrical current \cite{1511.08789},
%$\tau_{\chi}$ for CME current \cite{1608.08595}, $\tau_{D}$ for charge diffusion \cite{1511.08789} and the last $\tau_{\bar\chi}$ for the generalised CME. Relaxation terms are crucial in closing the formalism of MIS prescription. The relaxation term $\tau_{\chi}$ was recently computed numerically in \cite{Li:2018srq} for a similar holographic model but in the beyond probe limit.

Our second order results  could be compared  with similar results obtained in CKT \cite{1603.03442,1603.03620}.
%However, some manipulations are needed before the comparison. The first issue is on the off-shell nature of our formalism in deriving the currents.
To this goal, we first put the currents on-shell eliminating $\partial_t\rho$ and $\partial_t\rho_5$
using  the continuity equations (\ref{cont eqn}).
% to eliminate $\partial_t\rho$ and $\partial_t\rho_5$ terms in favour of spatial derivative terms.
Second, we replace the densities by the corresponding chemical potentials.
%The second issue is about the choice of hydrodynamical variables: density versus chemical potential.
In the holographic model, the  chemical potentials $\mu,\mu_5$ are computed analytically in the hydrodynamic limit. At second order in the gradient expansion,
\begin{align}\label{gen mu(2)}
\mu&=\frac{1}{2}\rho+\frac{1}{16}(\pi-2\log2)\vec{\nabla}^2\rho-\frac{3}{4}(\pi-2\log2)
\kappa (\vec{B}\cdot\vec{\nabla}\rho_5)+18(1-2\log2)\kappa^2\rho  B^2  \nonumber\\
&-\frac{1}{8}(\pi+2\log2)(\vec{\nabla}\cdot\vec{E})+\mathcal{O}(\partial^3),\\
%\end{align}
%%
%\begin{align}
\mu_5&=\frac{1}{2}\rho_5+\frac{1}{16}(\pi-2\log2)\vec{\nabla}^2\rho_5-\frac{3}{4}(\pi-2
\log2)\kappa (\vec{B}\cdot\vec{\nabla}\rho)+18(1-2\log2)\kappa^2 \rho_5 B^2  \nonumber\\
&+\frac{3}{2}(\pi-2\log2)\kappa (\vec{E}\cdot\vec{B}) +\mathcal{O}(\partial^3). \label{gen mu5(2)}
\end{align}
Eventually,  the on-shell currents are
\begin{eqnarray}\label{onj}
\vec{J}^{\rm on-shell}&=& \sigma_\chi^0 \mu_5 \vec{B}-\tau_{\chi}  \mu_5 \partial_t \vec{B}+\sigma_e^0 (\vec{E}-\vec{\nabla} \mu)-\tau_e\sigma_e^0 \partial_t \vec{E} +\sigma_{\chi H}^0\, \mu \vec{E}\times\vec{B}\\
&-&\mathcal{D}_H^0\vec{B}\times (\mu\vec{\nabla} \mu+\mu_5\vec{\nabla} \mu_5)- \sigma_{a\chi H}^0  \vec{E}\times \vec{\nabla}\mu_5 +\sigma_m^0 \left(\mu^2+\mu_5^2\right) \vec{\nabla}\times \vec{B} + \mathcal{O}(\partial^3), \nonumber
\end{eqnarray}
\begin{eqnarray}\label{onj5}
\vec{J}_5^{\rm on-shell}&=&\sigma_\chi^0 \mu \vec{B}-\tau_\chi \mu \partial_t\vec B-\sigma_e^0 \vec{\nabla} \mu_5 +\sigma_{\chi H}^0\, \mu_5 \vec{E}\times \vec B -\mathcal{D}_H^0 \vec{B} \times (\mu\vec{\nabla} \mu_5+\mu_5\vec{\nabla} \mu) \nonumber\\
&-&\sigma_{a \chi H}^0 \vec{E}\times \vec{\nabla}\mu +2\sigma_m^0 \mu\mu_5 \vec{\nabla}\times \vec{B} + \mathcal{O}(\partial^3).
\end{eqnarray}
Now lets discuss the physics of each term in (\ref{onj},\ref{onj5}), primarily focusing on $\vec{J}^{\,on-shell}$. The first  term in (\ref{onj}) is CME.
The next one introduces relaxation into CME
induced by time variation of the magnetic field, with $\tau_{\chi}$ being a relaxation time originally computed in \cite{1608.08595}.  $\tau_{\chi}$ was recently re-examined
numerically in \cite{Li:2018srq} within a quite similar holographic model but beyond probe limit.
The third and fourth terms are just the classic Ohm's and diffusion currents accompanied by another relaxation effect associated with time varying electric field.
The corresponding relaxation time $\tau_e$ was originally computed in \cite{Myers:2007we}.
Note that in (\ref{delJ 2nd},\ref{delJ5 2nd})  there are two additional relaxation time terms. The first one with $\tau_{D}$ enters the diffusion current \cite{1511.08789}.
Finally, $\tau_{\bar\chi}$ is yet another relaxation time associated with  generalised CME. Note the difference between $\tau_{\chi}$  and $\tau_{\bar\chi}$: while the former
is a TC responding to time varying external magnetic field, the latter is related to relaxation of the axial charge density.
In (\ref{onj},\ref{onj5}) both terms appear as $\mathcal{O}(\partial^3)$.

 $\sigma_{\chi}^0$ and $\tau_{\bar\chi}$ are the first two coefficients in the gradient expansion of a resummed TCF $\sigma_{\bar\chi}$  \cite{BDL}. Instead of a full resummation,
which is a complicated numerical problem, one could use the relaxation time $\tau_{\bar\chi}$ in order to build a causal model for $\sigma_{\bar\chi}$ in a spirit of MIS.
%will be collected into a TCF named $\sigma_{\bar\chi}$, which will play a crucial role in revealing the non-dissipative CMW mode \cite{BDL}.
In this sense, the relaxation times, such as  $\tau_\chi$ and $\tau_{\bar\chi}$, are  of special importance.
 %Both $\tau_\chi$ and $\tau_{\bar\chi}$ are important in formulating a causal hydrodynamical model in the sense of MIS prescription.

The $\vec{E}\times \vec B$-term in (\ref{onj},\ref{onj5}) looks very similar to the usual Hall effect, which  is, however,  absent in our holographic model because of the probe limit. The term that we do find is induced by the chiral anomaly. To distinguish it from the normal Hall effect,  it is  referred to as  {\it chiral Hall effect} \cite{1407.3168} with
$\sigma_{\chi H}^0$ being its TC. Notice that $\sigma_{\chi H}^0\propto \kappa^2$. Contrary to purely anomaly-induced effects, which are normally odd in $\kappa$, the terms even in $\kappa$ appear as anomaly-induced corrections to normal transports \cite{1105.6360}.
The $\mathcal{D}_H^0$-term generates current perpendicular to both the magnetic field and gradients of chemical potentials. In \cite{1603.03442} this effect was  called \textit{Hall diffusion}.
This term can be regarded as an example (we will expand on this below) in which the  diffusion constant is turned into a non-trivial  diffusion tensor depending on the magnetic field.

The $\sigma_{a\chi H}^0$-term in (\ref{onj}) induces flow  perpendicular to both  the electric field and gradient of the axial chemical potential.  It was referred to  as \textit{anomalous chiral Hall effect} in \cite{1603.03442}\footnote{Indeed, the $\tau_\chi$-,$\sigma_{\chi H}^0$-terms in (\ref{onj}) could be reorganised as $-\sigma_{a\chi H}^0(\vec E\times \vec\nabla \mu_5+\mu_5\partial_t \vec B)-\sigma_{a\chi H}^0 \mu_5 \partial_t \vec B$. More precisely it is  $(\vec E\times \vec\nabla \mu_5+\partial_t \vec B)$-term that was called anomalous chiral Hall effect in \cite{1603.03442}.}.

Finally, the last term in (\ref{onj}) corresponds to another anomaly-induced correction to a normal current.  Normal transport due to  rotor of magnetic field  was first analysed
in  \cite{1511.08789}. At  second order in the gradient expansion under discussion now, the ``normal'' transport coefficient  was found to be identically zero.
Thus, the entire effect at this order arises from the anomaly alone.

Dissipative nature of each term entering (\ref{onj},\ref{onj5}) is of interest. While we do not intend to dwell in this question here,
we note in passing that the   TCs  $\sigma_\chi^0$, $\tau_e$, $\sigma_{\chi H}^0$, $\mathcal{D}_H^0$, $\sigma_m^0$, and $\tau_D$  are all time reversal
$\mathcal{T}$-even and  thus non-dissipative.  The remaining terms in (\ref{onj},\ref{onj5}) are dissipative.

Starting from CSE, the various terms  in $\vec{J}_5^{\,on-shell}$ could be simply understood as axial analogues of those in $\vec{J}^{\,on-shell}$.

For the sake of a more detailed comparison of our results with parallel ones in CKT, we quote here the expression for the
vector current as appears  in \cite{1603.03442}
\begin{equation} \label{J ckt}
\begin{split}
\vec J_{\rm CKT}=&\frac{1}{2\pi^2}\mu_5 \vec B - \frac{\tau \mu_5}{6\pi^2} \partial_t \vec B+\sigma^{\rm CKT}_e(\vec E- \vec \nabla \mu)- \tau\sigma_e^{\rm CKT}\partial_t\vec E+ \sigma_H^{\rm CKT}\mu\vec E\times \vec B\\
&- \mathcal{D}_H^{\rm CKT}\left(\mu\vec\nabla \mu+ \mu_5\vec\nabla \mu_5\right)\times \vec B- \sigma_{a\chi H}^{\rm CKT}\vec E\times \vec \nabla \mu_5 - \mathcal{D}_{\chi}^{\rm CKT} \mu\mu_5 \vec\nabla \mu_5,
\end{split}
\end{equation}
where
\begin{equation}
\begin{split}
&\sigma_e^{\rm CKT}=\frac{\tau}{9\pi^2}\left[1+ 3 (\mu^2+ \mu_5^2) \right],\qquad  \sigma_H^{\rm CKT}=\frac{\tau^2}{3\pi^2},\qquad \mathcal{D}_H^{\rm CKT}= \frac{\tau^2}{3\pi^2},\\
&\sigma_{a\chi H}^{\rm CKT}= \frac{\tau}{6\pi^2},\qquad \mathcal{D}_{\chi}^{\rm CKT}= \frac{2\tau}{3\pi^2}.
\end{split}
\end{equation}
Here $\tau$ is a parameter of dimension of time introduced in relaxation time approximation (RTA) of CKT.
Confronting with (\ref{onj}) we notice absence of $\vec{\nabla}\mu_5$ term in $\vec J^{\ on-shell}$.
Similarly, there are no terms proportional to  $\vec{\nabla}\mu$, $\vec{E}$, $\partial_t \vec{E}$ in $\vec{J}_5^{\ on-shell}$. All these terms are expected to arise
beyond the probe limit.
On the other hand, the magnetic conductivity term $\vec{\nabla}\times \vec{B}$ is missing in (\ref{J ckt}).
%We remark that this normal transport receives anomalous corrections and thus becomes non-vanish \cite{1511.08789}.
All the remaining terms appear in perfect agreement, at least as far as general structures are concerned.

Because in principle the two models describe  two different
regimes (strong vs weak coupling), the transport coefficients are not expected to agree. It is nevertheless instructive to pursue such a comparison.
For this goal,  we need to fix the parameter $\tau$ of the CKT. Obviously, there is no unique way to fix $\tau$. We chose to set CME as a benchmark.
That is, we equate the CME conductivities and the associated relaxation times in two models. This results in
\begin{equation}
\kappa=\frac{1}{24\pi^2},\qquad \qquad \tau=3\log2.
\end{equation}
Then, the transport coefficients in (\ref{onj}) are
%\begin{equation}
%\begin{split}
%&\sigma_e^0=1,\qquad \tau_e\sim \mathcal{O}(10^{-1}),\qquad \sigma_{\chi H}^0\sim \mathcal{O} (10^{-4}), \qquad \mathcal{D}_H^0\sim \mathcal{O} (10^{-4}) \\
%&\sigma_{a\chi H}^0\sim \mathcal{O}(10^{-2}), \qquad \sigma_m^0\sim \mathcal{O}(10^{-4}),
%\end{split}
%\end{equation}
compared to those in (\ref{J ckt})
%\begin{equation}
%\begin{split}
%&\sigma_e^{\rm CKT}\sim \mathcal{O}(10^{-2}),\qquad \tau\sim \mathcal{O}(1),\qquad \sigma_H^{\rm CKT}=\mathcal{D}_H^{\rm CKT} \sim \mathcal{O}(10^{-1}),\\
%&\sigma_{a\chi H}^{\rm CKT}\sim \mathcal{O}(10^{-2}),\qquad \mathcal{D}_\chi^{\rm CKT} \sim \mathcal{O}(10^{-1}).
%\end{split}
%\end{equation}
\begin{equation}\label{compCKT}
\begin{split}
&\sigma_e^0/\sigma_e^{\rm CKT}\sim \mathcal{O}(10^{2}), \qquad \tau_e/\tau^{CKT}\sim\mathcal{O}(10^{-1}), \qquad \sigma_{\chi H}^0/\sigma_H^{\rm CKT}\sim\mathcal{O}(10^{-3}), \\
&\mathcal{D}_H^0/\mathcal{D}_H^{\rm CKT} \sim \mathcal{O}(10^{-3}), \qquad \sigma_{a\chi H}^0/\sigma_{a\chi H}^{\rm CKT}\sim \mathcal{O}(1).
\end{split}
\end{equation}
While some of the coefficients came out to be of the same order, the electrical conductivity $\sigma_e^{\rm CKT}$ in CKT is strongly suppressed (by order $10^{-2}$) compared to the holographic
model. On the other hand, the anomaly-induced coefficients
%Obviously, in contrast to weakly coupled results obtained within chiral kinetic theory, the electrical conductivity $\sigma_e^0$ is strongly enhanced (by order $10^2$) while $
$\sigma_{\chi H}^0$ and $\mathcal{D}_H^0$ are highly suppressed  (by order $10^{-3}$) in holography \footnote{
In the presented comparison, we have expressed the constitutive relations in terms of the chemical potentials, so to have them in  the same form as in  \cite{1603.03442}.
We could have done oppositely and that is to compare the results expressed in terms of the charge densities. To this goal we could use
 the relation between the charge  densities and the chemical potentials of \cite{1603.03442}, which remarkably are quite similar to  (\ref{gen mu(2)}, \ref{gen mu5(2)}).
 While on both sides of the comparison the TCs get modified, we have checked that the ratios (\ref{compCKT}) remain intact.
%We could have implemented this comparison by re-writing relevant results of \cite{1603.03442} in terms of charge densities.
%In the high temperature limit (corresponding to probe limit of this work), except for the values of various coefficients, the relations between charge densities and chemical potentials derived in \cite{1603.03442} are the same as ours (\ref{gen mu(2)}, \ref{gen mu5(2)}). Thus, it does not matter if we use $\rho,\rho_5$ or $\mu,\mu_5$
}.

%In presence of background electromagnetic fields, the charge diffusion and other conductivities turn into tensors, as in MHD.
A complimentary way of looking at (\ref{formal current1},\ref{formal current2},\ref{delJ 2nd},\ref{delJ5 2nd}) is by separately collecting
 terms proportional to $\vec\nabla\rho$ and $\vec\nabla\rho_5$. All these terms constitute a diffusive current, which to the lowest order in the gradients is
%we obtain $E,B$-dependent diffusion tensor
\begin{equation}
J^i_{\rm diff}= -\mathcal{D}_{ij}^0\nabla_j \rho- (\mathcal{D}_{\chi}^0)_{ij} \nabla_j \rho_5,
\end{equation}
where
\begin{equation}
\mathcal{D}_{ij}^0=\frac{1}{4}(4\mathcal{D}_0\delta_{ij}+ \mathcal{D}_H^0 \epsilon_{ikj} B_k \rho),\qquad (\mathcal{D}_{\chi}^0)_{ij}= \frac{1}{4}(2\sigma_{a\chi H}^0 \epsilon_{ikj} E_k+ \mathcal{D}_H^0 \epsilon_{ikj} B_k \rho_5).
\end{equation}

Much like in MHD, the diffusion constants are turned into  tensors, which in fact depend non-linearly on the external e/m fields $E$ and $B$.
%Here $\mathcal{D}_{ij}^0$ does not show dependence on external electric field.
Furthermore, when higher order gradients are resummed, these diffusion tensors  become momenta  dependent  tensor functions \cite{BDL}.
\\

\subsubsection{The third order results and collective excitations}

Third order corrections in $\delta \vec{J}$ and $\delta \vec{J}_5$ contain a few dozens of new terms with corresponding TCs, all of which are computed analytically
(see Section \ref{s3} for more details). Since the complete set of the results for $\delta \vec{J}^{\;[3]}$ and $\delta \vec{J}_5^{\;[3]}$ is very large,
in this Summary section we focus only on the most interesting terms, the ones which are  linear in $\rho,\rho_5$,
while the remaining nonlinear in $\rho,\rho_5$ corrections are flashed in Section \ref{s3}.

Denoted as $\delta \vec{J}^{\;[3]\,l}$ and $\delta \vec{J}_{5}^{\;[3]\,l}$, the linear in $\rho,\rho_5$ terms at third order are
%The terms that are linear in the densities are also the ones that contribute to the gapless waves  propagating in the chiral medium, such as CMW.  The dispersion relation will be derived below. Eventually, the linear in the charge densities parts of $\delta \vec{J}^{\;[3]}$ and $\delta \vec{J}_5^{\;[3]}$, denoted as $\delta \vec{J}^{\;[3]\,l}$ and $\delta \vec{J}_{5}^{\;[3]\,l}$, are
\begin{eqnarray}\label{delJ 3rd}
\delta \vec{J}^{\;[3]\,l}&&= \tau_1 \partial_t^2 (\vec{\nabla} \rho) +\tau_2 \vec{\nabla}^2 (\vec{\nabla} \rho) +\tau_3 \partial_t^2 \vec{E}+\tau_4 \vec{\nabla}^2 \vec{E}+\tau_5\partial_t(\vec{\nabla}\times\vec{B}) +\tau_6 (\partial_t^2\rho_5) \vec B \nonumber\\
&&+ \tau_7 \partial_t \rho_5 \partial_t \vec B+ \tau_8 \rho_5\partial_t^2 \vec B +\tau_9 (\vec{\nabla}^2\rho_5)\vec B+\tau_{10}\rho_5 \vec{\nabla}^2 \vec B +\tau_{11}\left(\vec \nabla\rho_5 \cdot \vec \nabla\right)\vec B\nonumber\\
&&+ \tau_{12} \left(\vec B\cdot \vec\nabla\right) \vec\nabla \rho_5 + \widetilde{\tau}_{12} \vec\nabla \left(\vec B\cdot \vec \nabla \right) \rho_5 +\underline{\tau_{13} \vec{\nabla}(\rho B^2)} +\underline{\tau_{14}\vec{E}\times (\vec{E} \times \vec{\nabla}\rho)} \nonumber\\
&&+\tau_{15} \partial_t (\vec{E} \times \vec{\nabla} \rho_5) + \tau_{16} \partial_t \vec{E}\times\vec{\nabla}\rho_5 + \tau_{17} \rho \vec{E} \times \partial_t \vec B+ \tau_{18} \partial_t (\rho \vec{B}\times\vec{E})   \nonumber\\
&&+ \tau_{19} \rho_5 \vec{E} \times(\vec{E}\times\vec{B})+\tau_{20} \vec{E}\times \partial_t (\rho\vec{B}),
\end{eqnarray}
\begin{eqnarray}\label{delJ5 3rd}
\delta \vec{J}_5^{\;[3]\,l}&&
= \tau_1 \partial_t^2 (\vec{\nabla} \rho_5) +\tau_2 \vec{\nabla}^2 (\vec{\nabla} \rho_5) +\tau_6 (\partial_t^2\rho) \vec B + \tau_7 \partial_t \rho \partial_t \vec B+ \tau_8 \rho\partial_t^2 \vec B +\tau_9 (\vec{\nabla}^2\rho)\vec B \nonumber\\
&& +\tau_{10}\rho \vec{\nabla}^2 \vec B+\tau_{11}\left(\vec \nabla\rho \cdot \vec \nabla \right)\vec B+ \tau_{12} \left(\vec B \cdot \vec\nabla\right) \vec\nabla \rho + \widetilde{\tau}_{12} \vec\nabla \left(\vec B\cdot \vec \nabla \right) \rho \nonumber\\
&&  +\underline{\tau_{13} \vec{\nabla}(\rho_5 B^2)}+ \underline{\tau_{14} \vec{E} \times (\vec{E}\times\vec{\nabla}\rho_5)} +\tau_{15} \partial_t (\vec{E} \times \vec{\nabla} \rho) + \tau_{16} \partial_t \vec{E}\times\vec{\nabla}\rho\nonumber\\
&&  + \tau_{17} \rho_5 \vec{E} \times \partial_t \vec B+ \tau_{18} \partial_t (\rho_5 \vec{B}\times\vec{E}) + \tau_{19} \rho \vec{E} \times(\vec{E}\times\vec{B})  +\tau_{20} \vec{E}\times \partial_t(\rho_5\vec{B})\nonumber\\
&& +\tau_{21}\vec{E}\times(\vec{\nabla}\times\vec{B}) +\tau_{22}\vec{\nabla} (\vec{B}\cdot\vec{E}) ,
\end{eqnarray}
where
\begin{equation}
\widetilde{\tau}_{12}=\tau_{12}+\tau_{10}-\tau_{9}.
\end{equation}
In (\ref{delJ 3rd}, \ref{delJ5 3rd}) we have made use of the Bianchi identity (\ref{Bianchi bdry}) and eliminated $\vec\nabla\times \vec E$.
The values of TCs $\tau_{1-22}$ are collected in Appendix \ref{appendixa1}, see (\ref{tau1}-\ref{tau22}). Apart from the $\tau_3,\tau_4,\tau_5,\tau_{21},\tau_{22}$ terms, one can obtain $\delta \vec{J}_5^{\;[3]\,l}$ from $\delta \vec{J}^{\;[3]\,l}$ via  exchange of $\rho$ and $\rho_5$. It is important to give physical interpretation for  $\tau_i$ in (\ref{delJ 3rd},\ref{delJ5 3rd}).

The TCs $\tau_{1-5}$ represent the second order gradient expansion of the charge diffusion function $\mathcal{D}$, electric and magnetic conductivity functions $\sigma_e,\sigma_m$, and were first computed in \cite{1511.08789} by employing weak field approximation. The $\tau_8,\tau_{10}$-terms are second order gradient expansion of CME conductivity $\sigma_\chi$ \cite{1608.08595}. The $\tau_{19}$-term  was first obtained in \cite{1608.08595} for constant electromagnetic fields, which, once expanded, contains nonlinear corrections to the original CME/CSE.

The underlined terms $\tau_{13},\tau_{14}$ include anomaly-induced $B^2$-, $E^2$-corrections to the charge diffusion constant $\mathcal{D}_0$:
\begin{equation}\label{d0c}
\mathcal{D}_0=\frac{1}{2}-18(2\log2-1)\kappa^2B^2-\frac{3}{4}\pi^2\kappa^2 E^2 +\cdots.
\end{equation}
We note that both corrections are negative, see (\ref{d0c}).  $E^2$-correction is new whereas $B^2$-correction was first calculated in \cite{1609.09054}. Obviously, there will be higher powers in $E^2,B^2$ corrections to $\mathcal{D}_0$. In the forthcoming publication \cite{Part3}, we will compute the charge diffusion constant, as a function of constant e/m fields relaxing the weak field approximation.

The transport coefficients $\tau_6,\tau_7,\tau_9$ are due to spacetime inhomogeneity of $\rho,\rho_5$. $\tau_6,\tau_9$ correspond to second order expansion of the generalised CME/CSE conductivity function $\sigma_{\bar{\chi}}$ to be computed in the forthcoming paper \cite{BDL}.

The terms $\tau_{11},\tau_{12},\widetilde{\tau}_{12}$ represent mixing effect between magnetic field and spatial gradients of $\rho,\rho_5$. They were first considered in \cite{1609.09054}. The TCs $\tau_{15-18},\tau_{20}$ have similar structure as the Hall diffusion and Hall effect, but the former are induced by time-varying densities and electromagnetic fields. Thus, the $\tau_{15-18},\tau_{20}$-terms are relaxation times corrections to the Hall diffusion and Hall effect.
%Indeed, they are crucial in formulating a causal hydrodynamical model if one wants to take into account these second order nonlinear effects in (\ref{delJ 2nd}, \ref{delJ5 2nd}).

The $\tau_{21},\tau_{22}$-terms are due to spatial inhomogeneity of electromagnetic fields. Vector analogs of $\tau_{21},\tau_{22}$ will emerge as nonlinear in $\rho,\rho_5$ terms, see $\tau_{30},\tau_{34}$-terms in (\ref{nonlinrhorho5j}).

Via the criterion for dissipative/non-dissipative transports based on $\mathcal{T}$-symmetry arguments, the TCs $\tau_{6-12}$, $\widetilde{\tau}_{12}$, $\tau_{15}$, $\tau_{16}$,$\tau_{19}$, $\tau_{21}$ and $\tau_{22}$ are $\mathcal{T}$-even and thus correspond to non-dissipative TCs, while the rest of the terms are dissipative.

The
third order gradient corrections (\ref{delJ 3rd}, \ref{delJ5 3rd}) contribute to various collective excitations of  the holographic chiral medium, particularly they
modify the dispersion relation of CMW \cite{1012.6026}. For constant electromagnetic fields,
\begin{eqnarray}\label{disp}
\omega=&&\pm\left[1-36(2\log2-1)\kappa^2 \mathbf{B}^2 -\frac{3\pi^2}{2} \kappa^2 \mathbf{E}^2 \right]6\kappa (\vec{q}\cdot\vec{\mathbf{B}}) \pm 9 \pi^2 (\vec{\mathbf{E}}\cdot\vec{\mathbf{B}})\kappa^3(\vec{q}\cdot\vec{\mathbf{E}})\nonumber\\
&&+(36\log2)\kappa^2(\vec{q}\cdot\vec{\mathbf{S}})-\left[\frac{1}{2}+18(1-2\log2)\kappa^2
\mathbf{B}^2-\frac{3\pi^2}{4}\kappa^2\mathbf{E}^2\right]i q^2\\
&&\pm \frac{9}{2}\log2 \kappa (\vec{q}\cdot\vec{\mathbf{B}})q^2-\frac{i}{8}q^4\log2 -i \frac{3}{4}\pi^2\kappa^2(\vec{q}\cdot\vec{\mathbf{E}})^2+i(36\log2) \kappa^2 (\vec{q}\cdot\vec{\mathbf{B}})^2+\cdots. \nonumber
\end{eqnarray}
When $\vec{\bf E}=0$, the dispersion relation (\ref{disp}) reduces to the one obtained in \cite{1609.09054}. The first term ($\sim \vec q\cdot \vec{\bf B}$) in (\ref{disp}) represents nonlinear corrections to the speed of CMW, which are negative making  the wave to propagate slower. The second term ($\sim \vec{q}\cdot\vec{\bf E}$) in (\ref{disp}) corresponds to a wave mode propagating along the electric field.  It is called  {\it density wave} \cite{Huang:2015oca} or {\it chiral electric wave} \cite{1407.3168}. Since the chiral electric separation effect vanishes in the probe limit, here this  effect is mimicked by the second term in (\ref{disp}) which is induced by the chiral anomaly as a nonlinear correction. Its presence is conditional to  $\vec{\bf E}$ not being orthogonal to $\vec{\bf B}$. We find the third term ($\sim \vec q\cdot \vec{\bf S}$)  of special interest because it corresponds to a new phenomenon.  It corresponds to a wave propagating along the direction of the energy flux $\vec{\bf S}=\vec{\bf E}\times \vec{\bf B}$, which can be  referred to as {\it chiral Hall density wave} (CHDW). The remaining terms in (\ref{disp}) are
decay rates of various wave modes.

The rest of this paper is structured as follows. Section \ref{s2} is about the holographic model. Section \ref{s3} is devoted to the main part of our study. Section \ref{s5} contains some closing remarks. Appendix \ref{appendixa1} collects more technical details. \\

\section{Holographic setup: $U(1)_V\times U(1)_A$}\label{s2}

The holographic model is Maxwell-Chern-Simons theory in the Schwarzschild-$AdS_5$. The bulk action is
\begin{equation}
S=\int d^5x \sqrt{-g}\mathcal{L}+S_{\textrm{c.t.}},
\end{equation}
where
\begin{equation}\label{LPVA}
\begin{split}
\mathcal{L}=&-\frac{1}{4} (F^V)_{MN} (F^V)^{MN}-\frac{1}{4} (F^a)_{MN} (F^a)^{MN} +\frac{\kappa\,\epsilon^{MNPQR}}{2\sqrt{-g}}\\
&\times\left[3 A_M (F^V)_{NP} (F^V)_{QR} + A_M (F^a)_{NP}(F^a)_{QR}\right],
\end{split}
\end{equation}
and the counter-term action $S_{\textrm{c.t.}}$ is
\begin{equation}\label{ct VA}
S_{\textrm{c.t.}}=\frac{1}{4}\log r \int d^4x \sqrt{-\gamma}\left[(F^V)_{\mu\nu} (F^V)^{\mu\nu} +(F^a)_{\mu\nu}(F^a)^{\mu\nu}\right].
\end{equation}
The gauge Chern-Simons terms ($\sim \kappa$) in the bulk action mimic the chiral anomaly of the boundary field theory.
Note $\epsilon^{MNPQR}$ is the Levi-Civita symbol with the convention $\epsilon^{rtxyz}=+1$, while the Levi-Civita tensor is $\epsilon^{MNPQR}/\sqrt{-g}$. The counter-term action (\ref{ct VA}) is specified based on minimal subtraction, which excludes {\it finite} contribution to the boundary currents from the counter-term.

In the ingoing Eddington-Finkelstein coordinate, the Schwarzschild-$AdS_5$ is
\begin{equation}
ds^2=g_{_{MN}}dx^Mdx^N=2dtdr-r^2f(r)dt^2+r^2\delta_{ij}dx^idx^j,
\end{equation}
where $f(r)=1-1/r^4$. Thus, the Hawking temperature (identified as temperature of the boundary theory) is normalised to $\pi T=1$.
On the hypersurface $\Sigma$ of constant $r$, the induced metric $\gamma_{\mu\nu}$ is
\begin{equation}
ds^2|_{\Sigma}=\gamma_{\mu\nu}dx^\mu dx^\nu=-r^2f(r)dt^2+r^2\delta_{ij}dx^idx^j.
\end{equation}

It is convenient to split the bulk equations into dynamical and constraint components,
\begin{equation}\label{eom VAmu}
\textrm{dynamical~~equations}:\qquad \textrm{EV}^\mu=\textrm{EA}^\mu=0,
\end{equation}
\begin{equation}\label{eom VAr}
\textrm{constraint~~equations}:\qquad \textrm{EV}^r=\textrm{EA}^r=0,
\end{equation}
where
\begin{equation}\label{EV}
\textrm{EV}^M\equiv \nabla_N(F^V)^{NM}+\frac{3\kappa  \epsilon^{MNPQR}} {\sqrt{-g}} (F^a)_{NP} (F^V)_{QR},
\end{equation}
\begin{equation}\label{EA}
\textrm{EA}^M\equiv \nabla_N(F^a)^{NM} +\frac{3\kappa \epsilon^{MNPQR}} {2\sqrt{-g}} \left[(F^V)_{NP} (F^V)_{QR}+  (F^a)_{NP} (F^a)_{QR}\right].
\end{equation}

The boundary currents are defined as
\begin{equation} \label{current definition}
J^\mu\equiv \lim_{r\to\infty}\frac{\delta S}{\delta V_\mu},~~~~~~~~~~~~~
J^\mu_5\equiv \lim_{r\to\infty}\frac{\delta S}{\delta A_\mu},
\end{equation}
which, in terms of the bulk fields, are
\begin{equation}\label{j bct}
\begin{split}
&J^\mu=\lim_{r\to\infty}\sqrt{-\gamma}\,\left\{(F^V)^{\mu M}n_{_M}+ \frac{6\kappa \epsilon^{M\mu NQR}}{\sqrt{-g}}n_{_M} A_N (F^V)_{QR}- \widetilde{\nabla}_\nu (F^V)^{\nu\mu}\log r \right\},\\
&J_5^\mu=\lim_{r\to\infty}\sqrt{-\gamma}\, \left\{(F^a)^{\mu M}n_{_M}+ \frac{2\kappa \epsilon^{M\mu NQR}}{\sqrt{-g}}n_{_M} A_N (F^a)_{QR}- \widetilde{\nabla}_\nu (F^a)^{\nu\mu}\log r \right\},
\end{split}
\end{equation}
where $n_{_M}$ is the outpointing unit normal vector with respect to the slice $\Sigma$, and $\widetilde{\nabla}$ is compatible with the induced metric $\gamma_{\mu\nu}$.

The radial gauge $V_r=A_r=0$ will be assumed throughout this work. As a result, in order to determine the boundary currents (\ref{j bct}) it is sufficient to solve dynamical equations (\ref{eom VAmu}) only, leaving the constraints aside. Indeed, the constraint equations (\ref{eom VAr}) give rise to continuity equations (\ref{cont eqn})
\begin{equation}\label{cont eq}
\partial_\mu J^\mu=0,~~~~~~~~~~~~~~~\partial_\mu J^\mu_5= 12\kappa \vec{E}\cdot \vec{B}.
\end{equation}
In this way, the currents' constitutive relations to be derived below are off-shell.

Practically, it is more instructive to relate the currents (\ref{j bct}) to the coefficients of near boundary asymptotic expansion of the bulk gauge fields. Near $r=\infty$,
\begin{equation}\label{asmp cov1}
V_\mu=\mathcal{V}_\mu + \frac{V_\mu^{(1)}}{r}+ \frac{V_\mu^{(2)}}{r^2}- \frac{2V_\mu^{\textrm{L}}}{r^2} \log r+\mathcal{O}\left(\frac{\log r}{r^3}\right),~~~~~~~~
A_\mu=%\mathcal{A}_\mu + \frac{A_\mu^{(1)}}{r}+
\frac{A_\mu^{(2)}}{r^2}%- \frac{2A_\mu^{\textrm{L}}}{r^2} \log r
+\mathcal{O}\left(\frac{\log r}{r^3}\right),
\end{equation}
where
\begin{equation}\label{asmp cov2}
V_\mu^{(1)}=\mathcal{F}_{t\mu}^V,~~~~~~~~~~~4V_\mu^{\textrm{L}}=\partial^\nu \mathcal{F}_{\mu\nu}^V.
\end{equation}
A possible constant term for $A_\mu$ in (\ref{asmp cov1}) has been set to zero, in accordance with the fact that no axial external fields is assumed to be present in the current study.
$\mathcal{V}_\mu$ is the gauge potential of external electromagnetic fields $\vec {E}$ and $\vec{B}$,
\begin{equation}
E_i=\mathcal{F}_{it}^V=\partial_i\mathcal{V}_t-\partial_t \mathcal{V}_i,~~~~~~~~~~~~ B_i=\frac{1}{2}\epsilon_{ijk}\mathcal{F}_{jk}^V=\epsilon_{ijk}\partial_{j}\mathcal{V}_k.
\end{equation}
Dynamical equations (\ref{eom VAmu}) are sufficient to derive (\ref{asmp cov1},\ref{asmp cov2}), where the near-boundary data $V_\mu^{(2)}$ and $A_\mu^{(2)}$ have to be determined by completely solving (\ref{eom VAmu}) from the horizon to the boundary. The currents (\ref{j bct}) become
\begin{equation}\label{bdry currents}
\begin{split}
J^{\mu}	=\eta^{\mu\nu}(2V_{\nu}^{(2)}+2V^{\textrm{L}}_{\nu}+\eta^{\sigma t} \partial_{\sigma} \mathcal{F}_{t\nu}^V),\qquad \qquad
J_{5}^{\mu}= \eta^{\mu\nu}2A_{\nu}^{(2)}.
\end{split}
\end{equation}

As the remainder of this section, we outline the strategy for deriving the constitutive relations for $J^\mu$ and $J_5^\mu$. To this end, we turn on finite vector/axial charge densities for the dual field theory, which are also  exposed to external electromagnetic fields.
Holographically, the charge densities and external fields are encoded in asymptotic behaviors of the bulk gauge fields. In the bulk, we will solve the dynamical equations (\ref{eom VAmu}) assuming the charge densities and external fields as given, but without specifying them explicitly.

Following \cite{1511.08789} we start with the most general static and homogeneous profiles for the bulk gauge fields satisfying the dynamical equations (\ref{eom VAmu}),
\begin{equation}\label{homogeneous solution}
V_\mu=\mathcal{V}_\mu-\frac{\rho}{2r^2}\delta_{\mu t},~~~~~~~~~~~~
A_\mu=-\frac{\rho_{_5}}{2r^2}\delta_{\mu t},
\end{equation}
where $\mathcal{V}_\mu,\rho,\rho_{_5}$ are all constants for the moment. Regularity at $r=1$ has been used to fix one integration constant for each $V_i$ and $A_i$. As explained below (\ref{asmp cov2}), the constant term in $A_\mu$ is set to zero. Through (\ref{bdry currents}), the boundary currents are
\begin{equation}
J^t=\rho,~~~J^i=0;~~~~~~~~~~~J_5^t=\rho_{_5},~~~J_5^i=0.
\end{equation}
Hence, $\rho$ and $\rho_{_5}$ are identified as the vector/axial charge densities.

Next, following the idea of fluid/gravity correspondence \cite{0712.2456}, we promote $\mathcal{V}_\mu,\rho,\rho_{_5}$ into arbitrary functions of the boundary coordinates
\begin{equation}
\begin{split}
\mathcal{V}_\mu\to \mathcal{V}_\mu(x_\alpha),~~~~~~~~~~~\rho \to \rho(x_\alpha), ~~~~~~~~~~~
%\mathcal{A}_\mu \to \mathcal{A}_\mu(x_\alpha),~~
\rho_{_5}\to \rho_{_5}(x_\alpha).
\end{split}
\end{equation}
As a result, (\ref{homogeneous solution}) ceases to solve the dynamical equations (\ref{eom VAmu}). To have them satisfied, suitable corrections  in $V_\mu$ and $A_\mu$ have to be introduced:
\begin{equation} \label{corrections}
\begin{split}
V_\mu(r,x_\alpha)=\mathcal{V}_\mu(x_\alpha)-\frac{\rho(x_\alpha)}{2r^2}\delta_{\mu t}+ \mathbb{V}_\mu(r,x_\alpha),\qquad
A_\mu(r,x_\alpha)=%\mathcal{A}_\mu(x_\alpha)+
-\frac{\rho_{_5}(x_\alpha)}{2r^2}\delta_{\mu t} + \mathbb{A}_\mu(r,x_\alpha),
\end{split}
\end{equation}
where $\mathbb{V}_\mu,\mathbb{A}_\mu$ will be determined by solving (\ref{eom VAmu}). Appropriate boundary conditions are classified into three types. First, $\mathbb{V}_\mu$ and $\mathbb{A}_\mu$ are regular over the domain $r\in [1,\infty)$. Second, at the conformal boundary $r=\infty$, we require
\begin{equation}\label{AdS constraint}
\mathbb{V}_\mu\to 0,~~~~~~\mathbb{A}_\mu \to 0~~~~~~~\textrm{as}~~~~~~r\to \infty,
\end{equation}
which amounts to fixing external gauge potentials to be $\mathcal{V}_\mu$ and zero (for the axial fields).
Additional integration constants will be fixed by the Landau frame convention for the currents,
\begin{equation}\label{Landau frame}
J^t=\rho(x_\alpha),~~~~~~~~~~~J^t_5=\rho_{_5}(x_\alpha).
\end{equation}
The Landau frame convention corresponds to a residual gauge fixing for the bulk fields.

The vector/axial chemical potentials are defined as
\begin{equation} \label{def potentials}
\begin{split}
\mu&=V_t(r=\infty)-V_t(r=1)=\frac{1}{2}\rho-\mathbb{V}_t(r=1),\\
\mu_{_5}&=A_t(r=\infty)-A_t(r=1)=\frac{1}{2}\rho_{_5}-\mathbb{A}_t(r=1).
\end{split}
\end{equation}
Generically, $\mu,\mu_{_5}$ are nonlinear functionals of densities and external fields.

In terms of $\mathbb{V}_\mu$ and $\mathbb{A}_\mu$, the dynamical equations (\ref{eom VAmu}) are
\begin{equation}\label{eom Vt}
0=r^3\partial_r^2 \mathbb{V}_t+3r^2 \partial_r \mathbb{V}_t+r\partial_r \partial_k \mathbb{V}_k+ 12\kappa \epsilon^{ijk} \left[\partial_r \mathbb{A}_i\left( \partial_j \mathcal{V}_k + \partial_j \mathbb{V}_k\right)+ \partial_r\mathbb{V}_i \partial_j \mathbb{A}_k\right],
\end{equation}
\begin{equation}\label{eom Vi}
\begin{split}
0&=(r^5-r)\partial_r^2 \mathbb{V}_i+(3r^4+1)\partial_r \mathbb{V}_i+2r^3\partial_r \partial_t \mathbb{V}_i-r^3 \partial_r\partial_i \mathbb{V}_t+ r^2\left(\partial_t \mathbb{V}_i- \partial_i \mathbb{V}_t\right)\\
&+r(\partial^2 \mathbb{V}_i - \partial_i\partial_k\mathbb{V}_k)-\frac{1}{2}\partial_i \rho +r^2\left(\partial_t\mathcal{V}_i-\partial_i\mathcal{V}_t\right)+ r\left(\partial^2 \mathcal{V}_i- \partial_i \partial_k \mathcal{V}_k\right)\\
&+12\kappa r^2\epsilon^{ijk}\left(\frac{1}{r^3}\rho_{_5}\partial_j \mathcal{V}_k +\frac{1}{r^3}\rho_{_5}\partial_j\mathbb{V}_k +\partial_r \mathbb{A}_t \partial_j \mathcal{V}_k+\partial_r \mathbb{A}_t \partial_j \mathbb{V}_k\right)\\
&-12\kappa r^2\epsilon^{ijk} \partial_r\mathbb{A}_j\left[\left(\partial_t \mathcal{V}_k- \partial_k \mathcal{V}_t\right)+\left(\partial_t \mathbb{V}_k- \partial_k \mathbb{V}_t\right)+\frac{1}{2r^2}\partial_k \rho\right]\\
&-12\kappa r^2\epsilon^{ijk}\left\{ \partial_r\mathbb{V}_j \left[\left(\partial_t \mathbb{A}_k- \partial_k \mathbb{A}_t\right)+\frac{1}{2r^2}\partial_k \rho_{_5}\right]-\partial_j \mathbb{A}_k\left(\partial_r \mathbb{V}_t +\frac{1}{r^3}\rho \right)\right\},
\end{split}
\end{equation}
\begin{equation}\label{eom At}
0=r^3\partial_r^2 \mathbb{A}_t+ 3r^2 \partial_r \mathbb{A}_t+r\partial_r \partial_k \mathbb{A}_k+ 12\kappa \epsilon^{ijk}\left[ \partial_r \mathbb{V}_i\left( \partial_j \mathcal{V}_k+ \partial_j \mathbb{V}_k\right)+ \partial_r\mathbb{A}_i  \partial_j \mathbb{A}_k\right],
\end{equation}
\begin{equation}\label{eom Ai}
\begin{split}
0&=(r^5-r)\partial_r^2 \mathbb{A}_i+(3r^4+1)\partial_r \mathbb{A}_i+2r^3\partial_r \partial_t \mathbb{A}_i-r^3 \partial_r\partial_i \mathbb{A}_t+ r^2\left(\partial_t \mathbb{A}_i- \partial_i \mathbb{A}_t\right)\\
&+r(\partial^2 \mathbb{A}_i - \partial_i\partial_k\mathbb{A}_k)-\frac{1}{2}\partial_i \rho_{_5} +12\kappa r^2\epsilon^{ijk}\left(\partial_j \mathcal{V}_k +\partial_j \mathbb{V}_k \right) \left(\partial_r \mathbb{V}_t +\frac{1}{r^3}\rho\right) \\
&-12\kappa r^2\epsilon^{ijk} \partial_r\mathbb{V}_j\left[\left(\partial_t \mathcal{V}_k- \partial_k \mathcal{V}_t\right)+\left(\partial_t \mathbb{V}_k- \partial_k \mathbb{V}_t\right)+\frac{1}{2r^2}\partial_k \rho\right]\\
&-12\kappa r^2\epsilon^{ijk}\left\{ \partial_r\mathbb{A}_j \left[\left(\partial_t \mathbb{A}_k- \partial_k \mathbb{A}_t\right)+\frac{1}{2r^2}\partial_k \rho_{_5}\right]- \partial_j \mathbb{A}_k \left(\partial_r \mathbb{A}_t +\frac{1}{r^3}\rho_{_5} \right) \right\}.
\end{split}
\end{equation}
In the next section we will present solutions to (\ref{eom Vt}-\ref{eom Ai}) under approximation  discussed in the Introduction.

%\section{Generic structure of vector/axial currents}\label{s3}

\section{Nonlinear chiral transport}\label{s3}

In this section, we initially explore generic structure of the vector and axial currents (\ref{const}) as emerges within the holographic model of Section \ref{s2}. No assumptions will be made regarding the charge densities $\rho$, $\rho_5$ and external fields $\vec E,\vec B$. While we are not able to solve the dynamical equations (\ref{eom Vt}-\ref{eom Ai}) analytically, we can advance by rewriting them in integral forms and extract near-boundary asymptotic expansion for the corrections $\mathbb{V}_{\mu}$ and  $\mathbb{A}_{\mu}$. The procedure is rather tedious. Hence all the details are moved to  Appendix \ref{appendixa1}. Via (\ref{bdry currents}), the near-boundary asymptotic behaviors (\ref{s3nbe1}-\ref{s3nbe4}) yield the results (\ref{formal current1},\ref{formal current2}) with $\delta \vec{J}$ and $\delta \vec{J}_5$ formally given by (\ref{deltaJJ5}). As is clear  from (\ref{eqnG},\ref{eqnH}), $\delta \vec{J}$ and $\delta \vec{J}_5$ are composed of higher derivative terms involving  ${\vec E}, \vec{B}$ and $\rho,\rho_5$.

Now we continue with the  gradient expansion of $\delta \vec{J}$ and $\delta \vec{J}_5$. Within the hydrodynamic limit,  the dynamical equations (\ref{eom Vt}-\ref{eom Ai}) are solved perturbatively. Let us introduce a formal expansion parameter $\lambda$ by $\partial_\mu \to \lambda \partial_\mu$, which counts order of the gradient expansion.
Then, $\mathbb{V}_\mu$ and $\mathbb{A}_\mu$ could be expanded in powers of $\lambda$,
\begin{equation} \label{VA perturb expansion}
\mathbb{V}_\mu= \sum_{n=1}^\infty \lambda^n \mathbb{V}_\mu^{[n]}, ~~~~~~~~~~~~~~~~~~~~~~~~~~~~~
\mathbb{A}_\mu= \sum_{n=1}^\infty \lambda^n \mathbb{A}_\mu^{[n]}.
\end{equation}
We remind the reader that for this study, $\vec{E}$ and $\vec{B}$ are considered to be of $\mathcal{O}(\lambda^1)$. At each order in $\lambda$, $\mathbb{V}^{[n]}_{\mu}$ and $\mathbb{A}^{[n]}_{\mu}$ obey a system of ODEs, which could be analytically solved via direct integration over $r$. We list the results for $\mathbb{V}^{[n]}_{\mu}$ and $\mathbb{A}^{[n]}_{\mu}$ up to $n=2$ in (\ref{VtAt 1st}-\ref{Ai 2nd}).

Inserting the first order results (\ref{VtAt 1st}-\ref{Ai 1st}) into (\ref{eqnG}-\ref{deltaJJ5}) produces the second order results for $\delta \vec{J}$ and $\delta \vec{J}_5$, as summarised in (\ref{delJ 2nd},\ref{delJ5 2nd}).
%In order to compare with the relevant results in the literature \cite{1603.03442,1603.03620}, we need to replace the charge densities in (\ref{delJ 2nd},\ref{delJ5 2nd}) by the chemical potentials, defined (\ref{def potentials}).
The results (\ref{VtAt 1st},\ref{Vt 2nd},\ref{At 2nd}) also lead to the expressions for the chemical potentials, as summarised in (\ref{gen mu(2)},\ref{gen mu5(2)}).

With the second order corrections $\mathbb{V}_\mu^{[2]}$ and $\mathbb{A}_\mu^{[2]}$ (\ref{Vt 2nd}-\ref{Ai 2nd}), we obtain the third order results $\delta \vec{J}^{\;[3]}$ and $\delta \vec{J}_5^{\;[3]}$. However, nonlinearity makes such calculations rather involved and the number of various terms is very large.
For the sake of presentation, we have split the third order corrections into terms that are linear in either $\rho$ or $\rho_5$, and the rest.

The linear in the charge densities parts of $\delta \vec{J}^{\;[3]}$ and $\delta \vec{J}_5^{\;[3]}$, denoted as $\delta \vec{J}^{\;[3]\,l}$ and $\delta \vec{J}_{5}^{\;[3]\,l}$, are already presented through (\ref{delJ 3rd},\ref{delJ5 3rd}).
%
%In (\ref{delJ 3rd},\ref{delJ5 3rd}), we have made use of the Bianchi identity (\ref{Bianchi bdry}) and eliminated $\vec\nabla\times \vec E$.
%The values of TCs $\tau_{1-22}$ are collected in Appendix \ref{appendixa1}, see (\ref{tau1}-\ref{tau22}). Apart from the $\tau_3,\tau_4,\tau_5,\tau_{21},\tau_{22}$ terms, one can obtain $\delta \vec{J}_5^{\;[3]\,l}$ from $\delta \vec{J}^{\;[3]\,l}$ via  exchange of $\rho$ and $\rho_5$. It is important to give physical interpretation for  $\tau_i$ in (\ref{delJ 3rd},\ref{delJ5 3rd}).
%
These terms are the ones that contribute to the gapless waves  propagating in the chiral medium. We focus on the case with constant external fields only. Consider a plane wave ansatz for the charge densities
\begin{eqnarray}
\delta\rho=e^{-i(\omega t -\vec{q}\cdot\vec{x})}\delta\tilde{\rho}, \qquad \quad  \delta\rho_5=e^{-i(\omega t -\vec{q}\cdot\vec{x})}\delta\tilde{\rho}_5.
\end{eqnarray}
Then, the continuity equations (\ref{cont eqn}) with the constitutive relations (\ref{formal current1}, \ref{formal current2}, \ref{delJ 2nd}, \ref{delJ5 2nd}, \ref{delJ 3rd}, \ref{delJ5 3rd}) turn into
\begin{eqnarray}\label{diseqn}
a \delta\tilde{\rho}+b \delta\tilde{\rho}_5=0, \qquad \quad b \delta\tilde{\rho}+a \delta\tilde{\rho}_5=12 \kappa (\vec{\mathbf{E}}\cdot\vec{\mathbf{B}}).
\end{eqnarray}
The explicit expressions for $a$ and $b$ are
\begin{eqnarray} \label{a coef}
a&&=-i\omega+\frac{1}{2}q^2+18(1-2\log2)\kappa^2 q^2 \mathbf{B}^2 -\frac{3\pi^2}{4} \kappa^2 q^2 \mathbf{E}^2+9(\pi-2\log2)\kappa^2(\vec{q}\cdot\vec{\mathbf{B}})^2 \nonumber\\
&&+\frac{3\pi^2}{4}\kappa^2 (\vec{q}\cdot\vec{\mathbf{E}})^2 +i\frac{\pi}{8}\omega q^2-\frac{\pi^2}{48}\omega^2 q^2-\frac{1}{16}(\pi-2\log2)q^4+i36\log2 \kappa^2 (\vec{q}\cdot\vec{\mathbf{S}})\\
&&-(18\mathcal{C}+\frac{21\pi^2}{8})\kappa^2\omega(\vec{q}\cdot\vec{\mathbf{S}}),\nonumber
\end{eqnarray}
\begin{eqnarray} \label{b coef}
b&&=i6\kappa(\vec{q}\cdot\vec{\mathbf{B}})-i\frac{3}{4}(\pi-2\log2)\kappa q^2 (\vec{q}\cdot\vec{\mathbf{B}})+i 216(1-2\log2)\kappa^3 \mathbf{B}^2 (\vec{q}\cdot\vec{\mathbf{B}}) \nonumber \\
&&-\frac{3}{2}(\pi+2\log2)\kappa\omega(\vec{q}\cdot\vec{\mathbf{B}})-i\frac{1}{8}(24 \mathcal{C}+\pi^2+6\log^2 2)\kappa \omega^2 (\vec{q}\cdot\vec{\mathbf{B}})\\
&&-i\frac{3}{4}(\pi-2\log2)\kappa q^2 (\vec{q}\cdot\vec{\mathbf{B}}) +i9\pi^2\kappa^3 [(\vec{\mathbf{B}}\cdot\vec{\mathbf{E}})(\vec{q}\cdot\vec{\mathbf{E}})-\mathbf{E}^2
(\vec{q}\cdot\vec{\mathbf{B}})],\nonumber
\end{eqnarray}
where the Poynting vector $\vec{\mathbf{S}}=\vec{\mathbf{E}}\times\vec{\mathbf{B}}$. For $\omega,q\ll 1$, the dispersion equation (\ref{diseqn}) can be solved perturbatively, leading to the $\mathbf{B}/\mathbf{E}$-corrected dispersion relation (\ref{disp}).

Finally, we turn to terms that are nonlinear in the charge densities in the third order results $\delta \vec{J}^{\,[3]}$ and $\delta \vec{J}_5^{\;[3]}$. We denote them as $\delta \vec{J}^{\;[3]\,nl}$ and $\delta \vec{J}_5^{\;[3]\,nl}$:
\begin{eqnarray}\label{nonlinrhorho5j}
\delta \vec{J}^{\;[3]\,nl}&&= \tau_{23}(\rho^2+\rho_5^2)\partial_t\vec\nabla\times \vec B +\tau_{24}\rho_5(\rho_5^2+3\rho^2)\vec{\nabla}^2\vec B+ \tau_{25} \partial_t\vec{H} +\tau_{26} (\partial_t \rho_5\vec{\nabla}\rho_5+\partial_t \rho \vec{\nabla}\rho)\times \vec B \nonumber\\
&&+\tau_{27} (\rho_5\partial_t \vec{\nabla}\rho_5+\rho\partial_t \vec{\nabla}\rho) \times \vec B +\tau_{28} (\rho_5\partial_t \rho_5+\rho \partial_t\rho) \vec \nabla\times \vec B +\tau_{29} (\rho_5\vec{\nabla}\rho_5+\rho \vec{\nabla}\rho) \times \partial_t \vec B \nonumber\\
&& + \tau_{30}24\kappa \rho\rho_5 \vec{\nabla}\times \vec{S}+ \tau_{31} (\rho_5 \vec{\nabla}\rho+ \rho\vec{\nabla} \rho_5 ) \times \vec{S}+\tau_{32} (\vec{\nabla}\rho_5 \times \partial_t\vec{\nabla}\rho + \vec{\nabla}\rho\times \partial_t \vec{\nabla} \rho_5)\nonumber\\
&&+\tau_{33}2\rho\rho_5 \vec B\times \partial_t\vec{B}+ \tau_{34} 2\rho\rho_5 \vec{E}\times (\vec{\nabla}\times \vec{B}) +\tau_{35} [\rho_5\vec{\nabla}\times (\vec{E}\times \vec{\nabla} \rho_5)+(\rho_5\to\rho)]\nonumber\\
&&+\tau_{36}[\vec{\nabla}\rho\times(\vec{E}\times \vec{\nabla}\rho)+(\rho\to\rho_5)]+ \tau_{37} [2\rho\rho_5\vec{\nabla}\rho+ (\rho^2+\rho_5^2)\vec{\nabla}\rho_5]\times (\vec{\nabla}\times \vec B)\nonumber\\
&&+\tau_{38} (\rho_5\vec{\nabla}\times\vec{H}+ \rho\vec{\nabla} \times \vec{H}_a) + \tau_{39}(\vec\nabla\rho_5\times\vec{H} + \vec\nabla\rho \times \vec H_a )+ \tau_{40}
\vec E\times \vec{H}_a,
\end{eqnarray}
\begin{eqnarray}\label{nonlinrhorho5j5}
\delta \vec{J}_5^{\;[3]\,nl}&&= \tau_{23}2\rho\rho_5\partial_t\vec\nabla\times \vec B +\tau_{24}\rho(\rho^2+3\rho_5^2)\vec{\nabla}^2\vec B+ \tau_{25} \partial_t\vec{H}_a +\tau_{26} (\partial_t \rho\vec{\nabla}\rho_5+\partial_t \rho_5 \vec{\nabla}\rho)\times \vec B \nonumber\\
&&+\tau_{27} (\rho\partial_t \vec{\nabla}\rho_5+\rho_5\partial_t \vec{\nabla}\rho) \times \vec B +\tau_{28} (\rho\partial_t \rho_5+\rho_5 \partial_t\rho) \vec \nabla\times \vec B +\tau_{29} (\rho\vec{\nabla}\rho_5+\rho_5 \vec{\nabla}\rho) \times \partial_t \vec B \nonumber\\
&& + \tau_{30}12\kappa(\rho^2+\rho_5^2)\vec{\nabla}\times \vec{S}+ \tau_{31} (\rho \vec{\nabla}\rho+ \rho_5\vec{\nabla} \rho_5 ) \times \vec{S}+\tau_{32} (\vec{\nabla}\rho \times \partial_t\vec{\nabla}\rho + \vec{\nabla}\rho_5\times \partial_t \vec{\nabla} \rho_5)\nonumber\\
&&+\tau_{33}(\rho^2+\rho_5^2) \vec B\times \partial_t\vec{B}+ \tau_{34} (\rho^2+\rho_5^2) \vec{E}\times (\vec{\nabla}\times \vec{B}) +\tau_{35} [\rho\vec{\nabla}\times (\vec{E}\times \vec{\nabla} \rho_5)+(\rho_5\leftrightarrow \rho)]\nonumber\\
&&+\tau_{36}[\vec{\nabla}\rho_5\times(\vec{E}\times \vec{\nabla}\rho)+ (\rho\leftrightarrow\rho_5)]+ \tau_{37} [2\rho\rho_5\vec{\nabla}\rho_5+ (\rho^2+\rho_5^2)\vec{\nabla}\rho]\times (\vec{\nabla}\times \vec B)\nonumber\\
&&+\tau_{38} (\rho\vec{\nabla}\times\vec{H}+ \rho_5\vec{\nabla} \times \vec{H}_a) + \tau_{39}(\vec\nabla\rho\times\vec{H} + \vec\nabla\rho_5 \times \vec H_a )+ \tau_{40}
\vec E\times \vec{H},
\end{eqnarray}
where
\begin{equation}
\vec{S}=\vec E\times \vec B,\qquad \vec H= \vec B\times(\rho_5\vec\nabla\rho_5+ \rho \vec\nabla\rho), \qquad \vec H_a= \vec B\times(\rho\vec\nabla\rho_5+ \rho_5 \vec \nabla \rho).
\end{equation}
All $\tau_i$'s in (\ref{nonlinrhorho5j},\ref{nonlinrhorho5j5}) are computed analytically and the results are deposited in Appendix \ref{appendixa1}, see (\ref{tau23}-\ref{tau40}). Below we give simple explanation for each term in (\ref{nonlinrhorho5j},\ref{nonlinrhorho5j5}).

The TC $\tau_{23}$ corresponds to anomalous corrections to the relaxation term in the magnetic conductivity $\sigma_m$ of \cite{1511.08789,1608.08595}. The analytical result for $\tau_{23}$ was unknown in \cite{1608.08595}. The $\tau_{28}$-term  is a nonlinear correction to the magnetic current ($\sigma_m$-term of \cite{1511.08789}), and relies on time-varying densities. $\tau_{37}$ corresponds to a mixing effect between the charge diffusion and magnetic current. The TC $\tau_{30}$ is due to spatial inhomogeneity of e/m energy flux and is an analog of the magnetic conductivity.

The $\tau_{24}$-term stands for second order expansion of the CME conductivity $\sigma_\chi$ of \cite{1608.08595} and was first computed there.
$\tau_{25}$ is the relaxation term for the second order Hall diffusion current (see the $\mathcal{D}_H^0$-term  in (\ref{onj},\ref{onj5})). The  $\tau_{26},\tau_{27},\tau_{29}$-terms rely on the time inhomogeneity of charge densities or magnetic field and could be thought of as extension of the Hall diffusion current. The TC $\tau_{31}$ is related to the e/m energy flux and also generalises the Hall diffusion current.

$\tau_{32}$ is composed of spatial gradient of charge densities and corresponds to {\it nonlinear charge diffusion} process. The $\tau_{36},\tau_{39}$-terms are e/m field corrections to the nonlinear charge diffusions.  The last TC $\tau_{40}$ is a nonlinear in $E,B$ correction to the charge diffusions. The terms $\tau_{33},\tau_{34}$ are nonlinear in densities corrections to $\tau_{21},\tau_{22}$.

$\tau_{35}$ is the third order extension of the anomalous chiral Hall effect, i.e., $\sigma_{a\chi H}^0$-term in (\ref{onj},\ref{onj5}). In \cite{BDL}, we will perform a systematic resummation for certain transports and the term $\tau_{35}$ will be generalised into a TCF. Similarly, $\tau_{38}$ can be simply taken as the magnetic analogue of $\tau_{35}$ and will be extended to a TCF in  \cite{BDL}.

Finally, let us mention the dissipative nature for each term in the third order results (\ref{nonlinrhorho5j},\ref{nonlinrhorho5j5}). Via the criterion of $\mathcal{T}$-symmetry, the TCs $\tau_{24}$, $\tau_{30-34}$, $\tau_{37-40}$ are $\mathcal{T}$-even and are thus non-dissipative. The remaining terms are all dissipative.

\section{Conclusion}\label{s5}

In this work, we have continued exploration of nonlinear chiral anomaly-induced transport phenomena based
on  a holographic model with two $U(1)$ fields interacting via a Chern-Simons term.
For a finite temperature system, we constructed  off-shell constitutive relations for the vector/axial currents. A detailed report on our new results could be found in the Summary section. Here they are in brief:
\begin{itemize}
\item We demonstrated that both CME and CSE get corrected by higher derivative terms, see (\ref{formal current1},\ref{formal current2}). In the hydrodynamic limit,
we analytically calculated those gradient corrections up to third order. Comparison with the CKT was presented.
%see (\ref{delJ 2nd},\ref{delJ5 2nd},\ref{delJ 3rd},\ref{delJ5 3rd}) or (\ref{onj} \ref{onj}).
New third order results, particularly  (\ref{delJ 3rd},\ref{delJ5 3rd}), extend those that were initially considered in \cite{1608.08595,1609.09054} and reveal
novel  effects associated with time-dependence or inhomogeneities of the charge densities and external fields.

\item Among new results worth highlighting,  in weak field approximation
the charge diffusion constant $\mathcal{D}_0$ was found to receive negative anomaly-induced ${\bf E}^2$- and ${\bf B}^2$-corrections (\ref{d0c}).
It is very interesting to explore the dependence of $\mathcal{D}_0$ on the e/m fields beyond the weak field approximation, that is non-perturbatively.  Of particular
interest would be a strong field limit. We are pursuing this line  of study in the forthcoming paper \cite{Part3} (see also \cite{1012.6026} for similar study but in a different holographic model\footnote{In \cite{1012.6026} the effect of non-perturbative magnetic field on the speed of CMW and diffusion constant was induced by  nonlinear DBI action, which is quite different from the model in \cite{Part3}.}).

\item Another result we found to be of interest is that  the chiral medium is shown to support three types of collective modes: CMW (propagating along $\vec{\bf B}$),
CEW  propagating along  $\vec{\bf E}$,  and a new  one, chiral Hall density wave, propagating orthogonal to the other two, that is, along the energy flux $\vec{\bf E} \times \vec{\bf B}$.
\end{itemize}
%We have also computed analytically several different  relaxation times associated with the first and second order transport phenomena.

The follow up paper \cite{BDL} focuses on another set of approximations. Instead of considering a fixed order gradient expansion adopted here,
 we compute some TCFs in nonlinear chiral transport phenomena.  More specifically, the external electromagnetic fields are assumed to be constant and weak, while
the  charge densities are split into constant backgrounds and small inhomogeneous fluctuations.  The setup is similar to that of \cite{1608.08595}, but in  \cite{BDL}
as opposed to \cite{1608.08595}, gradient  resummation is performed for terms that are  linear both in the
charge density fluctuations  and external fields.

%The final results were summarised in (\ref{ji11},\ref{j5i11}), where the TCFs were first computed analytically in the hydrodynamic limit (section \ref{s42}) and then numerically up to large frequency/momentum (section \ref{s43}). Overall, the TCFs in (\ref{ji11},\ref{j5i11}) depend on spatial momentum weakly but show remarkable behavior as functions of frequency, namely damped oscillations towards vanishing asymptotically at $\omega\simeq 5$.

We have found a wealth of non-linear phenomena all induced entirely by the chiral anomaly. An important next step in deriving a full chiral MHD would be to abandon
the probe limit adopted in this paper and include the dynamics of a neutral flow as well.  This will bring into the picture additional effects such as thermoelectric conductivities, normal Hall current,
the chiral vortical effect \cite{0809.2488,0809.2596}, and some nonlinear effects discussed in \cite{1603.03620}.  We plan to address these in the future.

\appendix

\section{Supplement for Section \ref{s3}}\label{appendixa1}

In this Appendix, we collect all calculational details omitted in section \ref{s3}. Regarding  the general structure of the constitutive relations of the vector/axial currents,
we present the integral versions of the bulk dynamical equations and explore near boundary asymptotics. We further derive the gradient  expansion, and compute analytically
all TCs,  up to third order.

The dynamical equations (\ref{eom Vt}-\ref{eom Ai}) can be directly integrated over $r$, resulting in the following integral forms
\begin{equation}\label{s3nbe1}
\begin{split}
\mathbb{V}_t=&- \int_r^\infty \frac{dx}{x^3}\int_x^\infty dy \left\{y \partial_y \partial_k \left(\mathbb{V}_k+\frac{E_k}{y}\right)+12\kappa \epsilon^{ijk} \partial_y \mathbb{A}_i \left(\partial_j \mathcal{V}_k + \partial_j \mathbb{V}_k \right)\right.\\
&\left. +12\kappa \epsilon^{ijk}\partial_y \mathbb{V}_i \partial_j \mathbb{A}_k \right\}
+ \partial_k E_k \left(\frac{\log r}{2r^2}+ \frac{1}{4r^2}\right)\\
&\xrightarrow[]{ r\rightarrow\infty } \partial_k E_k \left(\frac{\log r}{2r^2}+ \frac{1}{4r^2}\right) +\mathcal{O}\left(\frac{\log r}{r^3}\right),
\end{split}
\end{equation}
\begin{equation}
\begin{split}
\mathbb{V}_i=&-\int_r^\infty \frac{xdx}{x^4-1}\left\{-\partial_t E_i \log x + \frac{x-1} {2x}\partial_i\rho +(x-1)E_i + \epsilon^{ijk} \partial_j B_k \log x -12 \kappa B_i \right.\\
&\left.~~~~~~~~~~~~~~~~~~~\times\left[\mu_5 +\mathbb{A}_i(x)-\frac{\rho_5}{2x^2}\right]+ G_i(x)\right\}\\
&\xrightarrow[]{ r\rightarrow\infty } \left(\frac{\log r}{2r^2}+ \frac{1}{4r^2}\right) \left(\partial_t E_i -\partial_k \mathcal{F}_{ik} \right) -\frac{1}{4r^2} \partial_i \rho +\left(-\frac{1}{r}+ \frac{1}{2r^2}\right)E_i + \frac{6\kappa \mu_5 B_i}{r^2}\\ &-\frac{G_i(x=\infty)}{2r^2} +\mathcal{O}\left(\frac{\log r}{r^3}\right),
\end{split}
\end{equation}
\begin{equation}
\begin{split}
\mathbb{A}_t=&-\int_r^\infty \frac{dx}{x^3}\int_x^\infty \left\{y\partial_y \partial_k \mathbb{A}_k(y) +12\kappa \epsilon^{ijk}\partial_y \mathbb{V}_i\left(\partial_j \mathcal{V}_k + \partial_j \mathbb{V}_k \right) +12\kappa \epsilon^{ijk} \partial_y \mathbb{A}_i \partial_j \mathbb{A}_k\right\}\\
&\xrightarrow[]{ r\rightarrow\infty } \mathcal{O}\left(\frac{1}{r^3}\right),
\end{split}
\end{equation}
\begin{equation}\label{s3nbe4}
\begin{split}
\mathbb{A}_i=&-\int_r^\infty \frac{xdx}{x^4-1} \left\{\frac{x-1}{2x} \partial_i \rho_5 - 12 \kappa B_i \left[\mu+ \mathbb{V}_t(x)- \frac{\rho}{2x^2}\right] + H_i(x)\right\}\\
&\xrightarrow[]{ r\rightarrow\infty } -\frac{1}{4r^2} \partial_i \rho_5 +\frac{6 \kappa B_i \mu}{r^2} -\frac{1}{2r^2} H_i(x=\infty)+\mathcal{O}\left(\frac{1}{r^3}\right),
\end{split}
\end{equation}
where $\mu$ and $\mu_5$ are the chemical potentials defined in (\ref{def potentials}). We have also provided asymptotic expansions near the boundary $r=\infty$. The functions $G_i(x)$ and $H_i(x)$ are
\begin{equation}\label{eqnG}
\begin{split}
G_i(x)&=-\int_1^x dy\left\{ 2y\partial_y\partial_t \left[\mathbb{V}_i(y) +\frac{E_i}{y} \right]+ \partial_t \left(\mathbb{V}_i(y)+\frac{E_i}{y}\right) -y\partial_y\partial_i \mathbb{V}_t -\partial_i\mathbb{V}_t \right. \\
&~~~~~~~~~~~~~~~~\left.+\frac{1}{y} (\partial^2\mathbb{V}_i-\partial_i\partial_k \mathbb{V}_k)  +12\kappa\epsilon^{ijk}\left[\frac{1}{y^3}\rho_5\partial_j \mathbb{V}_k +\partial_y\mathbb{A}_t\partial_j\mathbb{V}_k\right] \right.\\
&~~~~~~~~~~~~~~~~\left.-12\kappa\epsilon^{ijk} \partial_y\mathbb{A}_j \left[(\partial_t \mathbb{V}_k -\partial_k\mathbb{V}_t)+\frac{1}{2y^2} \partial_k\rho- E_k\right] \right.\\
&~~~~~~~~~~~~~~~~\left. -12\kappa\epsilon^{ijk}\partial_y\mathbb{V}_j\left[
(\partial_t\mathbb{A}_k-\partial_k\mathbb{A}_t)+\frac{1}{2y^2}\partial_k\rho_5\right]
\right.\\
&~~~~~~~~~~~~~~~~\left.+12\kappa\epsilon^{ijk}\left(\frac{1}{y^3}\rho\partial_j \mathbb{A}_k+\partial_y \mathbb{V}_t\partial_j\mathbb{A}_k\right)\right\},
\end{split}
\end{equation}
\begin{equation}\label{eqnH}
\begin{split}
H_i(x)=-\int_1^x dy&\left\{2y\partial_y\partial_t\mathbb{A}_i-y\partial_y\partial_i \mathbb{A}_t +(\partial_t\mathbb{A}_i-\partial_i\mathbb{A}_t)+\frac{1}{y} (\partial^2 \mathbb{A}_i-\partial_i\partial_k\mathbb{A}_k)+12\kappa\epsilon^{ijk} \right. \\
&~\times\left[\frac{\rho}{y^3}\partial_j\mathbb{V}_k+\partial_y \mathbb{V}_t \partial_j \mathbb{V}_k\right] -12\kappa\epsilon^{ijk}\partial_y \mathbb{V}_j \left[(\partial_t\mathbb{V}_k-\partial_k\mathbb{V}_t)+\frac{\partial_k\rho}{2y^2}-E_k \right]\\
&~\left. -12\kappa\epsilon^{ijk}\partial_y\mathbb{A}_j\left[(\partial_t\mathbb{A}_k -\partial_k\mathbb{A}_t)+\frac{1}{2y^2}\partial_k\rho_5\right]+12\kappa\epsilon^{ijk}
\left(\frac{1}{y^3}\rho_5\partial_j\mathbb{A}_k \right.\right.\\ &~\left.\left.+\partial_y\mathbb{A}_t\partial_j\mathbb{A}_k\right)\right\}.
\end{split}
\end{equation}
In deriving (\ref{s3nbe1}-\ref{s3nbe4}), all three types of the boundary conditions, as summarized in section \ref{s2}, were used to fix the integration constants. The formal solutions (\ref{s3nbe1}-\ref{s3nbe4}) give rise to the general results (\ref{formal current1},\ref{formal current2}) with $\delta J^i$ and $\delta J^i_5$ given as
\begin{equation}\label{deltaJJ5}
\delta J^i=\partial_t E_i -G_i(x=\infty),\qquad \qquad \qquad \delta J^i_5 = -H_i (x=\infty).
\end{equation}

For generic profiles of $\rho,\rho_5,\vec E, \vec B$, we are not able to compute $G_i(x=\infty)$ and $H_i(x=\infty)$ analytically. So, we employ the standard hydrodynamic limit and evaluate them up to  third order in the gradient expansion (\ref{VA perturb expansion}). Perturbative solutions for $\mathbb{V}_\mu$ and $\mathbb{A}_\mu$ are collected below.
At  first order, $n=1$,
\begin{equation}\label{VtAt 1st}
\mathbb{V}_t^{[1]}= \mathbb{A}_t^{[1]}=0,
\end{equation}
\begin{align} \label{Vi 1st}
\mathbb{V}_i^{[1]}&=f_1(r)\partial_i \rho + f_3(r) E_i+f_2(r) \rho_5 B_i ,
\end{align}
\begin{equation}\label{Ai 1st}
\mathbb{A}_i^{[1]}=f_1(r)\partial_i \rho_5 +f_2(r) \rho B_i ,
\end{equation}
where
\begin{align}
f_1(r)=\frac{1}{8}\left[\log{\frac{(1+r)^2}{1+r^2}}+2\arctan(r)-\pi\right],\qquad f_2(r)=3\kappa \log{\frac{1+r^2}{r^2}},
\end{align}
\begin{align}
f_3(r)=\frac{1}{4}\left[\log{\frac{1+r^2}{(1+r)^2}}+2\arctan(r)-\pi\right].
\end{align}

At  second order, $n=2$,
\begin{equation}\label{Vt 2nd}
\mathbb{V}_t^{[2]}=a_0 \partial_k E_k+ a_1 \left(-\frac{1}{2}\partial^2\rho+6\kappa B_k \partial_k \rho_5\right) +a_2 72\kappa^2 B^2\rho,
\end{equation}
\begin{equation}\label{At 2nd}
\mathbb{A}_t^{[2]}=a_1 \left(-\frac{1}{2}\partial^2\rho_5+6\kappa B_k \partial_k \rho -12\kappa \vec{E}\cdot \vec{B}\right) +a_2 72\kappa^2 B^2\rho_5,
\end{equation}
\begin{equation} \label{Vi 2nd}
\begin{split}
\mathbb{V}_i^{[2]}&=b_0 \epsilon^{ijk}\partial_jB_k+ b_1\partial_t \partial_i \rho+ b_2 \partial_t E_i +b_3 6\kappa\partial_t(\rho_5 B_i)+ b_4 3\kappa \rho_5 \epsilon^{ijk} \partial_j E_k +b_5 36\kappa^2\epsilon^{ijk} \\
&\times\left[-\left(\rho^2+\rho_5^2\right) \partial_j B_k + \rho_5 B_j \partial_k \rho_5 + \rho B_j \partial_k \rho\right] + b_6 6\kappa \epsilon^{ijk}\left[E_j \partial_k \rho_5 +12 \kappa \rho B_j E_k\right]\\
&- b_7 36\kappa^2 \epsilon^{ijk}\left(\rho B_j \partial_k \rho+ \rho_5 B_j \partial_k \rho_5 \right),
\end{split}
\end{equation}
\begin{equation}\label{Ai 2nd}
\begin{split}
\mathbb{A}_i^{[2]}&=b_1 \partial_t \partial_i \rho_5 +b_3 6 \kappa \partial_t(\rho B_i) + b_4 3\kappa \rho \epsilon^{ijk} \partial_j E_k + b_5 36 \kappa^2 \epsilon^{ijk} \left(-2\rho\rho_5 \partial_j B_k+ \rho B_j \partial_k \rho_5\right.\\
&\left.+ \rho_5 B_j \partial_k \rho\right)+ b_6 6\kappa \epsilon^{ijk} \left(E_j \partial_k \rho +12 \kappa \rho_5 B_j E_k\right)-b_7 36\kappa^2 \epsilon^{ijk} \left[\rho_5 B_j \partial_k \rho + \rho B_j \partial_k \rho_5\right],
\end{split}
\end{equation}
where
\begin{equation}
a_0=\frac{1+2\log r}{4r^2} + \int_r^\infty \frac{dx}{x^3} \int_x^\infty dy\frac{y^2+y+1} {y(y^2+1)(y+1)},
\end{equation}
\begin{equation}
a_1=\int_r^\infty \frac{dx}{x^3} \int_x^\infty \frac{ydy}{(y^2+1)(y+1)},
\end{equation}
\begin{equation}
a_2=\int_r^\infty \frac{dx}{x^3} \int_x^\infty \frac{dy}{y(y^2+1)},
\end{equation}
\begin{equation}
b_0=-\int_r^\infty \frac{xdx}{x^4-1} \int_1^x \frac{dy}{y},
\end{equation}
\begin{equation}
b_1=-\int_r^\infty \frac{xdx}{x^4-1} \int_1^x dy\left\{-\frac{y}{(y^2+1)(y+1)}- \frac{1}{8} \left[\log \frac{(1+y)^2}{1+y^2}+2\arctan(y)-\pi\right]\right\},
\end{equation}
\begin{equation}
b_2=-\int_r^\infty \frac{xdx}{x^4-1} \int_1^x dy \left\{-\frac{2y^2}{(y^2+1)(y+1)} -\frac{1}{4} \left[\log\frac{1+y^2}{(1+y)^2}+2\arctan(y)-\pi \right]\right\},
\end{equation}
\begin{equation}
b_3=-\int_r^\infty \frac{xdx}{x^4-1} \int_1^x dy \left\{\frac{2}{y^2+1}-\frac{1}{2} \log \frac{1+y^2}{y^2}\right\},
\end{equation}
\begin{equation}
b_4=-\int_r^\infty \frac{xdx}{x^4-1} \int_1^x dy \left\{-\frac{1}{y^3} \left[ \log \frac{1+y^2}{(1+y)^2} +2\arctan(y)-\pi\right]\right\},
\end{equation}
\begin{equation}
b_5=-\int_r^\infty \frac{xdx}{x^4-1} \int_1^x dy\frac{1}{y^3}\log\frac{1+y^2}{y^2},
\end{equation}
\begin{equation}
b_6=-\int_r^\infty \frac{xdx}{x^4-1} \int_1^x \frac{dy}{y(y^2+1)},
\end{equation}
\begin{equation}
b_7=-\int_r^\infty \frac{xdx}{x^4-1} \int_1^x \frac{dy}{y^3(y^2+1)}.
\end{equation}

Substituting the first order solutions (\ref{VtAt 1st},\ref{Vi 1st},\ref{Ai 1st}) into (\ref{eqnG}, \ref{eqnH}) generates the second order results (\ref{delJ 2nd},\ref{delJ5 2nd}).
The chemical potentials (\ref{gen mu(2)},\ref{gen mu5(2)}) are obtained similarly by
substituting  the results (\ref{VtAt 1st},\ref{Vt 2nd},\ref{At 2nd}) into (\ref{def potentials}).
Finally, the solutions (\ref{VtAt 1st}-\ref{Ai 2nd}) give rise to the third order corrections (\ref{delJ 3rd},\ref{delJ5 3rd}) with the transport coefficients $\tau_{1-40}$ as
\begin{align}\label{tau1}
\tau_1= \int_1^\infty dy\left[2y \partial_yb_1(y)+b_1(y)\right]=-\frac{\pi^2}{48},
\end{align}
\begin{align}
\tau_2=\int_1^\infty dy \frac{1}{2}\left[y\partial_y a_1(y)+a_1(y)\right]= -\frac{1}{16} \left(\pi-2\log2\right),
\end{align}
\begin{align}
\tau_3=\int_1^\infty dy \left[2y \partial_y b_2(y) +b_2(y)\right]= -\frac{\pi^2}{24},
\end{align}
\begin{align}
\tau_4=-\int_{1}^\infty dy \left[y\partial_y a_0(y)+ a_0(y)\right]= \frac{1}{8} \left(\pi +2\log 2\right),
\end{align}
\begin{align}
\tau_5=\int_1^\infty dy \left[\frac{f_3(y)}{y}+(y\partial_y+1) a_0(y) +(2y \partial_y+1) b_0(y) \right] = -\frac{1}{8} \left(\pi -\frac{\pi^2}{2} +2\log2\right),
\end{align}
\begin{align}
\tau_6=\int_1^\infty dy 6\kappa\left[2y\partial_y b_3(y)+b_3(y)\right]=\frac{1}{8} \kappa \left(24\mathcal{C}+\pi^2+ 6\log^22\right),
\end{align}
\begin{align}
\tau_7=\int_1^\infty dy 3\kappa (2y\partial_y+1)[4b_3(y)+b_4(y)] =9\kappa \mathcal{C}+ \frac{5}{16}\kappa \pi^2+ \frac{3}{2}\kappa \log^22,
\end{align}
\begin{align}
\tau_8=\int_1^\infty dy\left\{3\kappa (2y\partial_y+1)[2b_3(y)+b_4(y)] + 12\kappa \frac{b_2(y)}{y^3}\right\} =\kappa \left(6\mathcal{C}+ \frac{1}{4}\pi^2\right),
\end{align}
\begin{align}
\tau_9=\int_1^\infty dy \frac{f_2(y)}{y}= \frac{1}{8}\kappa \pi^2,
\end{align}
\begin{align}
\tau_{10}=\int_1^\infty \left[\frac{f_2(y)}{y}-12\kappa \frac{b_0(y)}{y^3}\right] = \frac{1}{4}\kappa \pi^2,
\end{align}
\begin{align}
\tau_{11}=\int_1^\infty \left[\frac{2f_2(y)}{y}+12\kappa \frac{\partial_y b_0(y)} {2y^2} \right]= \frac{5}{16} \kappa\pi^2,
\end{align}
\begin{align}
\tau_{12}=-\int_1^\infty\left[6\kappa\left[y\partial_y a_1(y)+a_1(y)\right]+ \frac{f_2(y)}{y}\right]=\frac{1}{8}\kappa \left(6\pi-12\log2-\pi^2\right),
\end{align}
\begin{align}
\tau_{13}=-\int_1^\infty 72\kappa^2 \left[y\partial_ya_2(y)+a_2(y) \right]= 18\kappa^2 \left(2\log2-1\right),
\end{align}
\begin{align}
\tau_{14}=-\int_1^\infty dy 72\kappa^2 \partial_y b_6(y)= -\frac{3}{4}\kappa^2\pi^2,
\end{align}
\begin{eqnarray}
\tau_{15}&&=\int_1^\infty \left\{6\kappa \left[2y\partial_y b_6(y) +b_6(y)\right] -12\kappa \left[f_1(y)\partial_y f_3(y)+ \partial_y b_1(y)\right]\right\}\\
&&=\frac{3}{2}\kappa \mathcal{C}+ \frac{5}{32}\kappa\pi^2,
\end{eqnarray}
\begin{eqnarray}
\tau_{16}&&= \int_1^\infty dy 12\kappa \left[\partial_y b_1(y) +\partial_y(f_1(y)  f_3(y))-\frac{\partial_y b_2(y)}{2y^2}\right]\nonumber\\
%&&=-\frac{1}{8}\kappa\left[12\mathcal{C}+\frac{\pi^2}{4}-6\log2(\pi+2\log2)\right],
&&=-\frac{1}{8}\kappa\left[12\mathcal{C}+\frac{\pi^2}{4}-6\log^22\right],
\end{eqnarray}
\begin{align}
\tau_{17}=-\int_1^\infty dy 36\kappa^2 \partial_y b_4(y)=-\frac{3}{8} \kappa^2 \left(48\mathcal{C}- \pi^2\right),
\end{align}
\begin{align}
\tau_{18}= \int_1^\infty dy \left[72\kappa^2 \left(2y \partial_y b_6(y) +b_6(y)\right)-12\kappa f_3(y) \partial_y f_2(y)\right] = \frac{3\pi^2}{2} \kappa^2,
\end{align}
\begin{align}
\tau_{19}=\int_1^\infty dy 864 \kappa^3 \partial_y b_6(y)= 9\kappa^3 \pi^2,
\end{align}
\begin{align}
\tau_{20}=- \int_1^\infty dy 12\kappa \left[6\kappa \partial_y b_3(y)+ \partial_y(f_2(y) f_3(y))\right] =-\frac{3}{8} \kappa^2 \left(48\mathcal{C} +3\pi^2\right),
\end{align}
\begin{align}
\tau_{21}=\int_1^\infty dy 12\kappa \partial_y b_0(y)= \frac{3}{8}\kappa \pi^2,
\end{align}
\begin{align} \label{tau22}
\tau_{22}=-2 \tau_{14}= -\frac{3}{2} \kappa \left(\pi-2\log2\right),
\end{align}
\begin{align} \label{tau23}
\tau_{23}=\int_1^\infty dy\left\{\frac{36\kappa^2}{y^3}[2b_3(y)+b_4(y)] -36\kappa^2 [2\partial_y b_5(y)+b_5(y)]\right\}=\frac{3}{2}\kappa^2(\pi^2-12\log2),
\end{align}
\begin{align}
\tau_{24}=\int_1^\infty dy 432\kappa^3\left[\frac{1}{y^3}b_5(y)\right]=- 108\kappa^3 (\,\log2-1\,)^2,
\end{align}
\begin{align}
\tau_{25}=&\int_1^\infty dy\left\{36\kappa^2(2\partial_y+1)[b_5(y)-b_7(y)]+12 \kappa [f_2(y)\partial_y f_1(y)-f_1(y)\partial_y f_2(y)]\right. \nonumber\\
&\qquad\qquad\left.-\frac{18\kappa^2}{y^3}[y\partial_yb_4(y)+2y\partial_yb_3(y)+4b_3(y) +2b_4(y)]\right\} \nonumber\\
=&\frac{3}{16}\kappa^2\left[-144\mathcal{C}+13\pi^2 +72\log^22 +12\pi (9\log2-4)\right],
\end{align}
\begin{align}
\tau_{26}&=-\int_1^\infty dy\left\{\frac{18\kappa^2}{y^3}[4b_3(y)+2b_4(y)+y\partial_y b_4(y)]+12\kappa f_1(y)\partial_y f_2(y)\right\} \nonumber\\
&=\frac{9}{4}\kappa^2\left[-8\mathcal{C}+(8+5\pi)\log2\right],
\end{align}
\begin{align}
\tau_{27}&=-\int_1^\infty dy\left\{\frac{18\kappa^2}{y^3}[2b_4(y)+2y\partial_yb_3(y) +y\partial_y b_4(y)] -12\kappa f_2(y)\partial_y f_1(y)\right\} \nonumber\\
&=\frac{9}{16}\kappa^2\left[48\mathcal{C}+\pi^2-4(8+7\pi)\log2\right],
\end{align}
\begin{align}
\tau_{28}=\int_1^\infty dy72\kappa^2\left[\frac{b_3(y)}{y^3}-[2\partial_y +1]b_5(y) \right] =\frac{3}{4}\kappa^2\left[\pi (5\pi-12)+12 (\pi-2) \log2\right],
\end{align}
\begin{align}
\tau_{29}=-\int_1^\infty dy 12\kappa f_1(y)\partial_y f_2(y) =- \frac{3}{16} \kappa^2 \left[48\mathcal{C}-\pi(\,\pi+12\log2\,)\right],
\end{align}
\begin{align}
\tau_{30}=-\int_1^\infty dy 72\kappa^2\left[\frac{1}{y^3}b_6(y)\right]= \frac{9}{2} \kappa^2(\log2)^2,
\end{align}
\begin{align}
\tau_{31}=-\int_1^\infty dy \frac{432\kappa^3}{y^3}\left[y\partial_yb_6(y)+2b_6(y) \right]= -\frac{9}{2}\kappa^3\left(\pi^2-24\log^22\right),
\end{align}
\begin{align}
\tau_{32}=\int_1^\infty dy 12\kappa\left[\frac{1}{2y^2}\partial_y b_1(y)-f_1(y) \partial_y f_1(y)\right]= -\frac{1}{64}\kappa\left[48\mathcal{C}+\pi(\,\pi-24\log2\,)\right],
\end{align}
\begin{align}
\tau_{33}=-\int_1^\infty dy 12\kappa f_2(y)\partial_y f_2(y) = 54\kappa^3(\log2)^2,
\end{align}
\begin{align}
\tau_{34}=\int_1^\infty dy 432\kappa^3 \partial_y b_5(y) = 9\kappa^3\left[\,\pi^2-6\log2(\,\log2-2\,)\right],
\end{align}
\begin{align}
\tau_{35}=\int_1^\infty dy 72\kappa^2\left[\frac{1}{y^3}b_6(y)\right]= -\frac{9}{2}\kappa^2(\log2)^2,
\end{align}
\begin{align}
\tau_{36}=\int_1^\infty dy 72\kappa^2 \left[\frac{1}{2y^2}\partial_y b_6(y)\right] = \frac{3}{8}\kappa^2\left[\,\pi^2-12(\,\log2\,)^2\right],
\end{align}
\begin{align}
\tau_{37}=-\int_1^\infty dy \frac{216\kappa^3}{y^3}\left[y\partial_yb_5(y)+4b_5(y) \right] =\frac{9}{2}\kappa^3\left[72-\pi^2+30(\log2-2)\log2\right],
\end{align}
\begin{align}
\tau_{38}=-\int_1^\infty dy 432\kappa^3\left[\frac{1}{y^3}b_7(y)-\frac{1}{y^3}b_5(y) \right]= \kappa^3\left(324-135\log^22+162\right),
\end{align}
\begin{align}
\tau_{39}=\int_1^\infty dy \frac{216\kappa^3}{y^2} \left[\partial_y b_5(y) -\partial_yb_7(y)\right] = -\frac{27}{4}\kappa^3\left[24-\pi^2-16\log2(\log2-2)\right],
\end{align}
\begin{align} \label{tau40}
\tau_{40}=-\int_1^\infty dy 432\kappa^3 \left[\partial_yb_5(y)-\partial_y b_7(y)\right] =-\frac{27}{2} \kappa^3\left[\,\pi^2-4\log2(\,\log2-4\,)\right],
\end{align}
where the Catalan's constant $\mathcal{C}\approx0.915966$.

\section*{Acknowledgements}

We would like to thank Umut G$\ddot{\rm u}$rsoy, Dmitri Kharzeev, Nathan Kleeorin, Shu Lin, Igor Rogachevskii, Andrey Sadofyev, Ho-Ung Yee for useful discussions. YB would like to thank the hospitality of the Department of Physics at  Ben-Gurion University of the Negev  where this work was initialised and finalised. YB was supported by the Fundamental Research Funds for the Central Universities under grant No.122050205032 and the Natural Science Foundation of China (NSFC) under the grant No.11705037.
TD and ML were supported by the Israeli Science Foundation (ISF) grant \#1635/16 and the BSF grants \#2012124 and \#2014707.

\providecommand{\href}[2]{#2}\begingroup\raggedright\endgroup

%\bibliographystyle{utphys}
%\bibliography{reference}

\end{document}